\documentclass[aps,twocolumn,secnumarabic,balancelastpage,amsmath,amssymb,nofootinbib,hyperref=pdftex]{revtex4}

\usepackage{ulem}           
\usepackage{soul}           
\usepackage{chapterbib}    
\usepackage{color}         
\usepackage{graphics}      
\usepackage[pdftex]{graphicx}      
\usepackage{longtable}     
\usepackage{epsf}          
\usepackage{bm}            
\usepackage{verbatim}			
\usepackage{listings}       

\usepackage[colorlinks=true]{hyperref}  
                                        
\usepackage{subfigure}

\newcommand{\lcdm}{\ensuremath{\Lambda\mathrm{CDM}}}

 \def\be{\begin{equation}}
\def\ee{\end{equation}}
\def\ba{\begin{eqnarray}}
\def\ea{\end{eqnarray}}
 
\begin{document}
\title{Quintessential Cosmological Tensions}
\author         {Arsalan Adil}
\email{aadil@ucdavis.edu}
\author{Andreas Albrecht}
\email          {ajalbrecht@ucdavis.edu}
\date{\today}
\affiliation{Center for Quantum Mathematics and Physics and Department of Physics and Astronomy\\ UC Davis, One Shields Ave, Davis CA.}
\author{Lloyd Knox}
\email          {lknox@ucdavis.edu}
\affiliation{Department of Physics and Astronomy\\ UC Davis, One Shields Ave, Davis CA.}

\begin{abstract}
Several cosmological tensions have emerged in light of recent data, most notably in the inferences of the parameters $H_0$ and $\sigma_8$. We explore the possibility of alleviating both these tensions \textit{simultaneously} by means of the Albrecht-Skordis ``quintessence'' potential. The field can reduce the size of the sound horizon $r_s^*$ while concurrently suppressing the power in matter density fluctuations before it comes to dominate the energy density budget today. Interestingly, this rich set of dynamics is governed entirely by one free parameter that is of $\mathcal{O}(10)$ in Planck units. We find that the inferred value of $H_0$ can be increased, while that of $\sigma_8$ can be decreased, both by $\approx 1\sigma$ compared to the \lcdm\ case. However, ultimately the model is disfavored by Planck and BAO data alone, compared to the standard $\Lambda$CDM model, with a $\Delta \chi^2 \approx +6$.  When including large scale structure and supernova data $\Delta \chi^2 \approx +1$.  We note that historically much attention has been focused on preserving the three angular scales $\theta_D$, $\theta_{EQ}$, and $\theta_s^*$ to their $\Lambda$CDM values. Our work presents an example of how, while doing so indeed maintains a relatively good fit to the CMB data for an increased number of ultra-relativistic species, it is a-priori insufficient in maintaining such a fit in more general model spaces.
\end{abstract}

\maketitle


\section{Introduction}

Much attention in cosmology today is devoted to tensions that exist between modern cosmological data and the standard ``$\Lambda$-Cold Dark Matter'' ($\Lambda$CDM) cosmological model. Extensions to $\Lambda$CDM which modify the matter content around the time of last scattering have been explored in the hopes of addressing these tensions. The Albrecht-Skordis (AS) model~\cite{Albrecht:1999rm,Skordis_2002} is a quintessence model in which the dark energy takes the form of an evolving scalar field with a number of special properties. One of these is that the scalar field contributes a significant fraction of the total energy density of the Universe starting at extremely early times, suggesting that it might offer relief from the cosmological tensions. 

This paper offers a systematic analysis of the AS model in the context of modern cosmological data.  We use the term ``ASCDM'' to refer to the model containing the same constituents as the $\Lambda$CDM model but where the cosmological constant $\Lambda$ has been replaced with the AS quintessence field for dark energy. We carefully study the features of the ASCDM model which suggest a resolution to the tensions, but ultimately show how these features are unable to make enough of a difference.   We find that the best fit ASCDM model fits the data overall about as well as $\Lambda$CDM.

Quintessence has been dismissed as a possible remedy to the tensions in the literature often on the basis of simple parameterizations \cite{impact-of-theoretical-priors} (e.g. the CPL parameterization \cite{CPL-1, CPL-2} or the $w$CDM parameterization where the dark energy equation of state is varied) or by employing ``tracking'' and/or ``freezing'' quintessence models as in \cite{Ooba-Ratra-Sugiyama, DIVALENTINO2019100385, H0-Assistance-Caldwell-Sabla}. Such approaches may artificially disfavor the potential of quintessence to alleviate the cosmological tensions: a low-order Taylor expansion is simply unable to capture the rich dynamics and early-vs-late behavioral changes that a quintessence field, such as that of the AS model, can exhibit\cite{Linden-CPL-mapping, Scherrer-CPL-mapping}; models that track the dominant energy component throughout cosmic history, and crucially \textit{during matter domination}, are also set up for failure in bringing concordance between the various measurements, as we explain in Sect.~\ref{sec:3} and Appendix A; finally, models where the field is ``frozen'' solely by means of Hubble friction, approach the $w_\phi \lesssim -1$ slowly \cite{Ooba-Ratra-Sugiyama, RP1988} and hence appear to be unfavorable candidates for resolving discrepancies in recent cosmological data (particularly the $H_0$ tension; discussed in Sect.~\ref{sec:2}). This paper avoids the pitfalls of these various approximation schemes by using the full evolution of the AS quintessence according to its field equations in our analysis. 

In addition to presenting the results of the constraints on model parameters given various data sets, we attempt to understand these constraints by comparing the ASCDM model with a much more well-studied model: the extension of \lcdm\ to allow for a variable number of light and dark degrees of freedom, parameterized by the effective number of neutrino species $N_{\rm eff}$ \cite[e.g.][]{dvorkin:2022}. We find that \textit{for similar total mean energy density at} $z > z_{\rm EQ}$, the best-fit ASCDM model is highly disfavored relative to the best-fit \lcdm\ + $N_{\rm eff}$ model. We speculate that this is fundamentally due to the changed shape of the ratio of the mean density of dark matter to the density of smooth components (radiation plus the AS field) as a function of scale factor, as this influences gravitational potential decay, which in turn impacts the amplitude of the acoustic standing wave oscillations as a function of wavelength \cite{Hu-White-CMB-Damping-Tail}. 

For context for this speculation, let us remind the reader that studies of constraints on light relics \cite[e.g.][]{hou2013,follin15} have emphasized the importance of three angular scales: those of the comoving horizon at matter-radiation equality, the comoving sound horizon, and the comoving photon diffusion scale, all projected from last scattering to today \cite{hu2001}. Keeping these angular scales fixed, in light relics models, as one increases the density of light relics, keeps one in or near regions of high CMB likelihood. They are very useful scales to keep in mind for understanding constraints on light relics, and for other model spaces as well. Others have emphasized their more general importance (\cite[e.g.][]{hubblehunters}) and used these scales to evaluate and understand results from early dark energy models \cite[e.g.][]{poulin2019}. 

The matter-radiation equality angular scale is important exactly because of its impact on gravitational potential decay and resulting impact on acoustic oscillation amplitude as a function of wavelength. Our results illustrate what may be a key point for model building: getting the angular scale right is not sufficient; the whole shape of what Hu \& White refer to as the ``radiation-driving envelope"~\cite{Hu-White-CMB-Damping-Tail} is important. This shape depends on how conditions at horizon crossing are changing as a function of scale factor, and therefore wavelength and angular scale.  The data are sensitive not just to a single scale, but to details in the decade to decade and a half of scale factor evolution prior to matter-radiation equality. 

The sensitivity of the CMB data to modes that are crossing the horizon during these times was also previously emphasized in \cite{hubblehunters}, as was the fact that in general a departure from \lcdm\ would result in a change in the shape of the radiation-driving envelope. We present here a specific illustration of this general principle. 

This paper is organized as follows:  Section~\ref{sec:2} presents an overview of the tensions and past attempts at a resolution. 
Section~\ref{sec:3} reviews the AS model, addressing both the background evolution and the evolution of perturbations. Section~\ref{sec:4} presents the elements of our analysis including data sets, parameters, priors and software tools used. Our results are presented in Sect.~\ref{sec:5}. Section~\ref{sec:5} also includes our systematic comparison with the ``\lcdm\ + $N_{\rm eff}$'' model which helps build intuition and explore the larger implications of our results.  In Sect.~\ref{sec:6} we reflect on the implications of what we have learned for other quintessence models, and we present our conclusions in Sect.~\ref{sec:7}.  Appendix~\ref{sec:A} briefly presents results from the ``Brane Model'' (a quintessence model similar to the AS model studied side-by-side with it in~\cite{Skordis_2002}) and draws some lessons about the unsuitability of pure ``tracking'' quintessence models.

\section{Cosmological tensions and attempts at resolution}
\label{sec:2}
Here we give a brief overview of the two cosmological ``tensions'' under discussion in cosmology today.
The first, and most prominent, of these is a tension in the measurement of the expansion rate of the Universe today, $H_0$, between cosmological- model-dependent and cosmological-model-independent (often referred to as ``early'' versus ``late'' Universe) probes. This discrepancy is most significant between the Cepheid calibrated supernovae method of the SH$_0$ES team, who find $H_0 = 73.2 \pm 1.04  \textnormal{ km/s/Mpc}$\cite{Riess:2021jrx} and the $\Lambda$CDM model calibrated with CMB data from the Planck satellite, which results in $H_0 = 67.4 \pm 0.5  \textnormal{km/s/Mpc}$ (using temperature, polarization, and lensing data)\cite{PCP18}. However, it would be mischaracterizing the seriousness of the issue were we to limit attention to just these experiments; indeed the discrepancy persists even if the Planck data are substituted, or complemented, with other CMB probes. For example, the inverse-distance ladder approach for calibrating supernovae data, which only depends on CMB data via a prior on the sound horizon scale, results in a discrepancy of up to $\approx3.9 \sigma$ with SH$_0$ES \cite{inv-dist-ladder, BOSS-EBOSS-inv-dist-ladder, Addison_2018-inv-dist-ladder}. One could also entirely neglect CMB data and calibrate $\Lambda$CDM using a combination of Big Bang Nucleosynthesis (BBN), Baryonic Accoustic Oscillations (BAO), and Large Scale Structure (LSS) data, which results in a value of $H_0$ discrepant at $\approx3.2-3.6 \sigma$ with SH$_0$ES \cite{BAO-BBN-LSS, BAO-BBN-2, BAO-BBN}. On the flip side, one could substitute or complement the SH$_0$ES data with other late-Universe probes, almost all of which continue to reflect the tension with varying levels of significance. We do note though that the supernova community has not all converged on acceptance of a ``high" value of the Hubble constant. In particular, Freedman et al. \cite{freedman:2019} find that the Tip of the Red Giant Branch technique for calibrating supernovae leads to distances to nearby galaxies that are significantly discrepant with the Cepheid-determined distances. They find a lower mean value for $H_0$ than the SH$_0$ES result, one that is consistent with the \lcdm\ Planck value (though also consistent at $\sim 2\sigma$ with SH$_0$ES as well.)    For an extensive review of the observational situation we refer the reader to \cite{Shah:2021}, and for an extensive review of the outcomes from various combinations of data we refer the reader to \cite{Verde19}, and to \cite{Riess19-2} for a more succinct collection of results. 

Due to the persistence of the tension across a plethora of data sets, it is natural to seek theoretical solutions to the problem. For an extensive compilation of the solutions see \cite{Valentino-review-solns} and for a quantitative comparison of them see \cite{h0olympics}. 

We consider the potential of alleviating the tension by means of an early-time modification to the expansion rate, which has some parallels with the Early Dark Energy (EDE), increased ultra-relativistic species, and Ratra-Peebles quintessence proposals (and other similar approaches), in that they all effectively increase the inference of $H_0$ from Planck by reducing the size of the sound horizon (discussed in detail in Section 3.1).  

There is also increasingly precise data available from cosmic shear and galaxy clustering surveys which independently constrain the weighted amplitude of the variance in matter fluctuations, $S_8 = \sigma_8  \sqrt{\Omega_m/0.3}$. These large-scale structure (LSS) surveys hint at a discrepancy with the Planck calibrated \lcdm\ model at a level of $\approx 2-4\sigma$.


However, the statistical significance of this second `tension' is more difficult to quantify for several reasons. First, the analysis has sensitivity to the choice of the two-point statistics, for example the ``COSEBIs'' approach of \cite{Asgari_2021} leads to $S_8 = 0.759^{+0.024}_{-0.026}$ while the traditional two-point correlation function (``2PCFs'') approach yields $S_8 = 0.764^{+0.018}_{-0.017}$ (68\% credible intervals). Second, there is a residual degeneracy that remains with $\Omega_m$ when fixing the exponent in the expression for $S_8$ so that changing the exponent may change the Gaussian difference measure. For example \cite{Asgari_2021} find a deviation of $3.4\sigma$, as opposed to $3\sigma$, from the Planck constraint when the exponent is varied.  Third, and related to the second point, is that the marginalized posteriors for the summary parameter have deviations from Gaussianity so that it is insufficient to characterize the tension by using a Gaussian difference measure. The statistically inclined reader is referred to Section 3.2 of \cite{Heymans_2021} (who find an $\approx3\sigma$ Gaussian discrepancy between Planck and KiDS-1000) and references therein for a discussion of various statistical measures that may be used to quantify the tension. Finally, it is non-trivial to combine data from different experiments which have overlapping regions of observation, although doing so leads to consistently lower values of the $S_8$ parameter than those inferred from Planck \cite{Heymans_2021}. Interestingly, combining cosmic shear data with anisotropic galaxy clustering is able to break the degeneracy in the $\sigma_8-\Omega_m$ plane, which indicates that the discrepancy is driven by a lower matter fluctuation amplitude, $\sigma_8$, as opposed to $\Omega_m$ \cite{Heymans_2021},\cite{cosmology-intertwined-III}. For a recent review of the status of this discrepancy and a compilation of results, we refer the reader to \cite{cosmology-intertwined-III}. A recent thorough review of the stats of both the aforementioned cosmological tensions, along with other statistical anomalies, can be found in \cite{Abdalla:2022yfr}.

Despite the challenges associated with quantifying the ``$S_8$ (or $\sigma_8$) tension'', it is possible that therein lies an important clue in the search for a concordant cosmological model. Certainly, a theory that resolves both these tensions simultaneously seems appealing and, therefore, in this work we make a point of trying to address the $S_8$ and $H_0$ tensions together. This position is not always taken in the literature: some models alleviate the discrepancy in $H_0$ but make that in $S_8$ more severe. 
For example, in the case of EDE, the inference of $H_0$ from Planck and BAO can be brought into concordance with the SH$_0$ES measurement but at the cost of an aggravated $\sigma_8$ discrepancy, so much so that the preference for EDE disappears on inclusion of the LSS data.\cite{EDE-Hill-LSS, EDE-Ivanov-LSS}. A similar narrative holds true for the case of an increased number of neutrino species, parameterized by $\Delta N_{\rm eff}$, as pointed out in Section 7.5.2 of \cite{PCP18}. Similarly, Ref.~\cite{Troster-S8-beyond-LCDM} studied a variety of extended-$\Lambda$CDM cosmologies, and found that the discrepancy persists in the extensions studied there, including for $w$CDM models where the discrepancy shifts from the marginalized $S_8$ posterior to the $S_8-w$ plane. In this article, we show that a background contribution from a quintessence field generically suppresses $\sigma_8$ and that the ASCDM model, in particular, is able to increase the inference of $H_0$ while simultaneously decreasing the inference of $\sigma_8$ from Planck data.


As we explain in Sect.~\ref{sec:5}, the perturbations to the scalar field also play a crucial role in driving the parameter inferences, and as such it is both theoretically and practically inconsistent to disregard the perturbations as is done in \cite{Roy-no-perturbations} and as the authors themselves point out in \cite{Roy-with-perturbations} (the implications of inconsistently accounting for the scalar field perturbations was first noted in  \cite{Caldwell-quintessence-perturbations-imprint}). The specific model we study here contributes significantly to the energy density prior to recombination, but less so after and appears to represent a best-case scenario for a single, minimally coupled scalar field with fairly generic initial conditions in alleviating the value of $H_0$ inferred from Planck. We also solve the equations of motion for the background and perturbations directly (Sect.~\ref{sec:5}). Still, we find that while such a model can relax the tensions, it is ultimately unsuccessful in bringing concordance between the various datasets. We leave the reader with some open questions regarding the origins of the tight constraints on quintessence as well as further avenues that we wish to explore with regards to these constraints. 

\section{Albrecht-Skordis model}
\label{sec:3}
The Albrecht-Skordis (AS) model, first introduced in~\cite{Albrecht:1999rm} studied extensively in~\cite{Skordis_2002} and further motivated in~\cite{Albrecht:2001cp,Albrecht:2001xt} is a particular quintessence model with a scalar field evolving in a potential of the form 
\begin{equation}
    V(\phi)=V_{0}\left((\phi-B)^{2}+A\right) e^{-\lambda \phi}.
    \label{eq:AS}
\end{equation}
This potential admits a local minimum due to the polynomial pre-factor. However, far from the minimum, the field behavior is dominated by the exponential factor so that it behaves similarly to an exponential potential, as introduced in~\cite{RP1988} and further studied in~\cite{FerJoyce}. At late late times the field may get ``stuck'' in the local minimum where it can approximate the behavior of a cosmological constant. 

In order to get the complete field behavior we solve the Friedmann equation, 
\begin{equation}
    H^2(a) = \left( \frac{\dot{a}}{a} \right)^2 = \frac{1}{3} \sum_i{\rho_i(a)}
    \label{eq:friedmann}
\end{equation}
simultaneously with the Klein-Gordon equation for scalar fields,
\begin{equation}
    \ddot{\phi} + 3 H\dot{\phi}+\partial_\phi V = 0
    \label{eq:KG}
\end{equation}
where we work in reduced Planck units with $m_p^{-2} = 8 \pi G/c^2 = 1$. The homogeneous pressure and energy density of the field are given the usual relationship,
\begin{equation}
\begin{split}
    \rho(\phi) = \frac{1}{2}\dot{\phi}^2 + V(\phi)
    \\
    p(\phi) = \frac{1}{2}\dot{\phi}^2 - V(\phi)
    \label{eq:density-pressure}
\end{split}
\end{equation}

\begin{figure}[p!t]
\centering
\includegraphics[width=9cm]{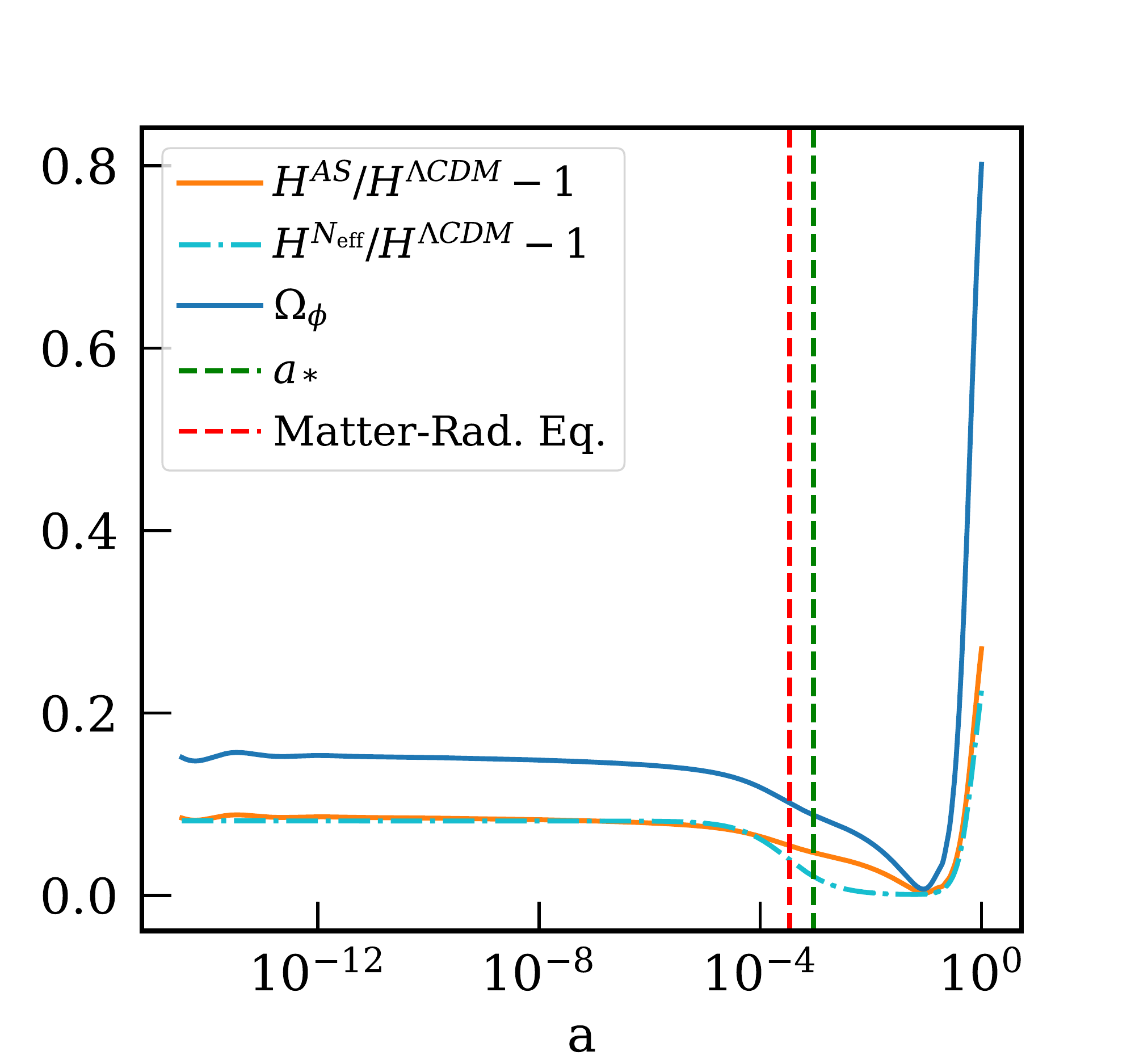}
\caption{The contribution to the energy density budget $\Omega_\phi$, and the fractional change in the Hubble parameter, $\delta H(a)/H(a)$, for the ASCDM model compared to $\Lambda$CDM using the Planck 2018 best-fit cosmological parameters and $\lambda=5$. The low value of $\lambda$ has been chosen only to exaggerate the behavior of the field. We also show the fractional change to the Hubble parameter for an increased $N_{\rm eff}$ model corresponding to the same early contribution as in the AS case ($N_{\rm eff}=4.32$).} 
\label{fig:delH}
\end{figure}

\begin{figure}[p!t]
\centering
\includegraphics[width=8.5cm]{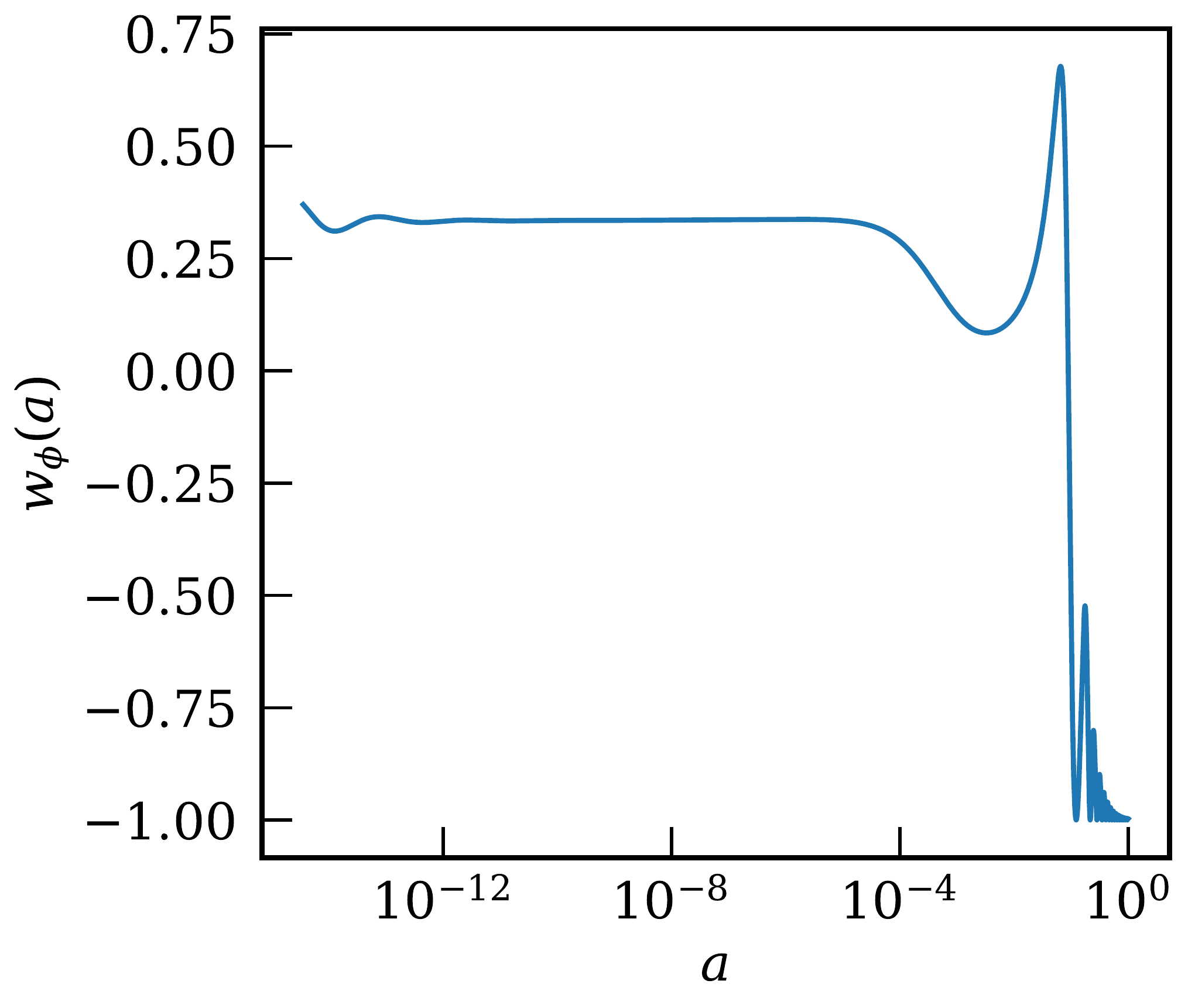}
\caption{After an initial transient, $w_\phi(a)$ tracks the dominant background component and breaks away from the tracker at the onset of matter domination before joining the new attractor near the minimum of the potential where it undergoes damped oscillations. } 
\label{fig:w-phi}
\end{figure}

The behavior of the field throughout cosmic history is summarized in Figs.~\ref{fig:delH} and \ref{fig:w-phi}. In Fig.~\ref{fig:w-phi} one can see that during radiation domination, the field has the same equation of state as the dominant background component. In this regime, the field is far from the minimum so that the behavior of the potential is dominated by the exponential factor in Eqn.~\ref{eq:AS} and we find, to a good approximation, that $\Omega^{\rm early}_\phi \approx 4/\lambda^2$ (see Fig.~\ref{fig:delH}). This is the same as the analytic ``tracking'' solution found by \cite{FerJoyce} for the pure-exponential potential, with the assumption that there is only a radiation component in addition to the scalar field. 
However, during matter domination, the field breaks away from this tracking behavior and $\Omega_\phi$ approaches zero. As we will see, this departure from tracking during matter domination is key for any hope of increasing $H_0$ beyond the $\Lambda$CDM value. Finally, the field lands in the local minimum of the potential so that, as one can see from Eqn.~\ref{eq:density-pressure}, $w_\phi\approx-1$ and the field quickly begins to dominate the total energy density (close to today). 

\subsection{Background Evolution}

In order to bring concordance between CMB and late-universe data one must take care not to disturb the angular scale of the sound horizon at recombination, $\theta_s^*$ since it is tightly constrained by Planck data \cite{hubblehunters}\footnote{This is a bit of a simplification since the constraint on $\theta_s^*$ fundamentally comes from Planck's sensitivity to the spacing between acoustic peaks and hence anything that impacts that spacing could change the inferred $\theta_s^*$. The temporal phase shift in acoustic oscillations from free-streaming neutrinos \cite{bashinsky04} is one such source of changes to peak spacing, but the impact is relatively small. See \cite{pan16} for more details on peak locations.} dependence to this constraint which fundamentally comes from the dependence of the spacing between peaks on $\theta_s^*$. Since

$$\theta_s^* = \frac{r_s^*}{D_M^*}$$
where 
\begin{equation}
r_{s}^{*}=\int_{z_{*}}^{\infty} \frac{d z}{H(z)} c_{s}(z)
\label{eq:rs}
\end{equation}
is the comoving size of the sound horizon at recombination and,
\begin{equation}
D_{M}^{\star}=\int_{0}^{z_{*}} \frac{d z}{H(z)} 
\label{eq:DM}
\end{equation}
is the comoving angular diamter distance to the last-scattering surface, an increased $H_0$ (at fixed $\omega_m$) would decrease $D_M^*$ and must therefore be compensated by a proportionate decrease in $r_s^*$.

To understand how the presence of the AS field can increase the inference of $H_0$, let us consider the case where all cosmological parameters of interest, and most importantly $\theta_s^*$ and $\omega_m$, are kept fixed to their Planck calibrated $\Lambda$CDM values but now with an added contribution from the scalar field (parameterized by $\lambda$). Since the field has a non-negligible contribution prior to recombination, the sound horizon decreases by an amount $\delta r_s^*/r_s \approx -4/(3\lambda^2)$ compared to the $\Lambda$CDM value (this would be $-2/\lambda^2$, but $\Omega_\phi$ begins to decrease at the onset of matter-radiation equality which softens this response). We can then approximate that $\frac{\delta H_0}{H_0} \approx \frac{20}{3\lambda^2}$ so that, for a $\approx 7\%$ increase in $H_0$ one requires $\lambda \approx 10$. This response in $H_0$, at fixed $\theta_s^*$ and $\omega_m$, is shown in Fig.~\ref{fig:delH} for the case of $\lambda=5$ which, from the approximation above, should yield $\delta H_0/H_0 \approx 0.27$. In Fig.~\ref{fig:delH} we also show the change in the expansion rate compared to $\Lambda$CDM for a model with an increased number of massless neutrinos. One can notice that, even though the background expansion of the ASCDM model is very similar to that of the \lcdm\ + $N_{\rm eff}$ model prior to matter-radiation equality, the response to $H_0$ is more efficient in the AS case for the same {\it early contribution} to $H(z)$ vs. the increased $N_{\rm eff}$ case. This efficiency is also important as there are independent constraints on the early expansion rate from measurements of light element abundances (see Section 7.6 in \cite{PCP18} and references therein) and thus further motivates us to study the implications of the ASCDM model for the $H_0$ tension. 

We note that the above exercise was done while keeping $\omega_m$ fixed. We will see in what follows that as the AS field energy density is increased, the best-fit model has an increased $\omega_m$ which qualitatively alters these conclusions.

\subsection{Perturbation Evolution}
\begin{figure}
\includegraphics[width=9cm]{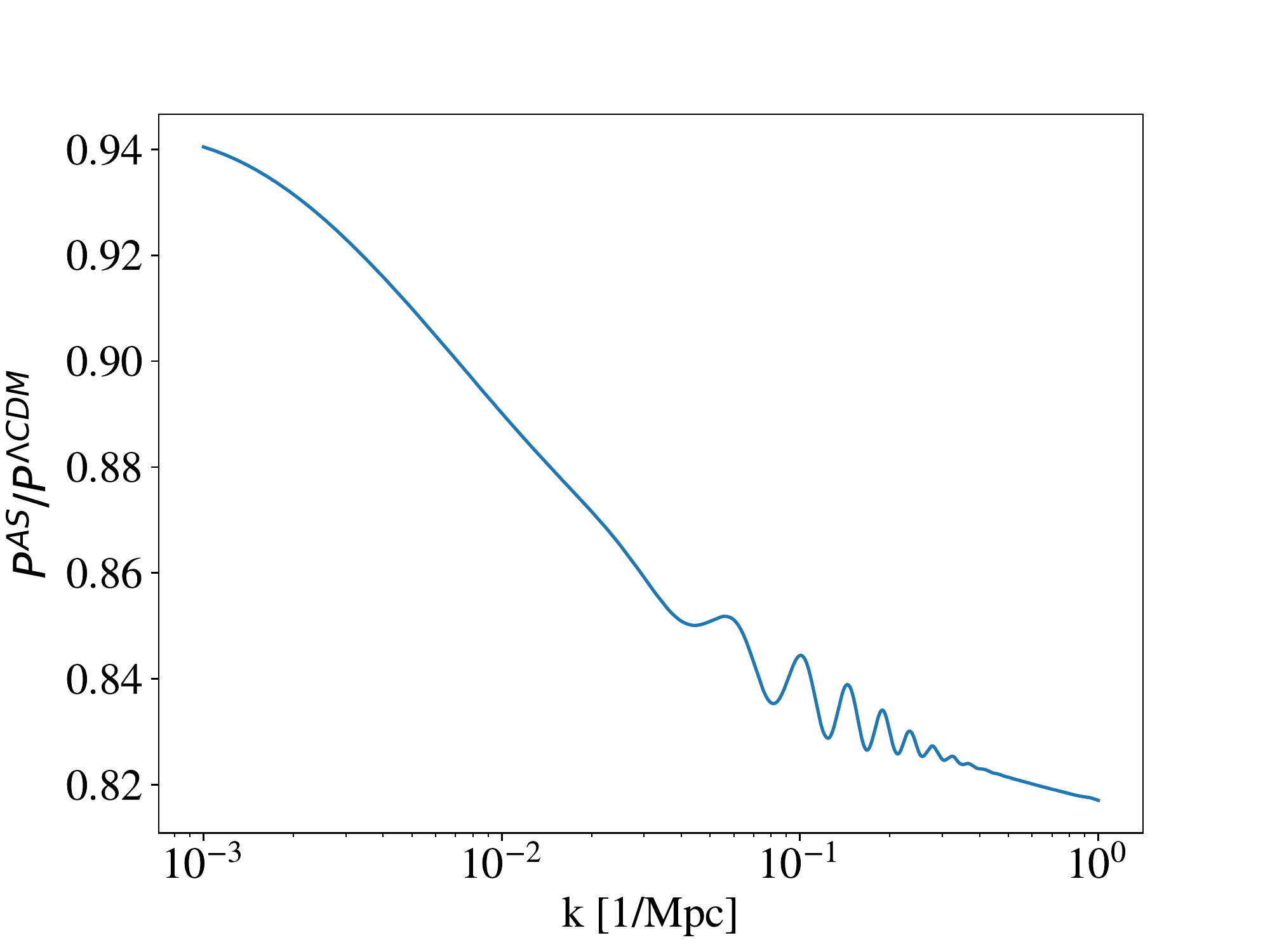}
\caption{For modes that enter before matter-radiation equality, there is a nearly scale-independent suppression (high-k regime). For modes that enter after, suppression is scale dependent since $\Omega_\phi$ begins to decrease at onset of matter domination.} 
\label{fig:mpk}
\end{figure}

We start by noting that the effective speed of sound for quintessence is $c_s^{\rm eff} = 1$ (in the gauge co-moving with the field) which prevents the component from clustering on sub-horizon scales \cite{Lesgourgues_and_Ballesteros}. In particular, even when $w_\phi \approx 0$, the entropy perturbations prevent gravitational collapse \cite{Skordis_2002}. Intuitively, this suggests that one should expect some power suppression since there is now an additional contribution to the expansion rate but a negligible contribution to gravitational potentials. 

Mathematically, notice that, late after horizon crossing, the evolution of the matter contrast $\delta_c$ is given by
\begin{equation}
\delta_c'' + aH\delta_c' = 4\pi a^2 \sum_i \rho_i \delta_i 
\label{eq:delta-m}
\end{equation}
(note the over-prime denotes derivative with respect to conformal time). In the standard cosmological model, one can find an approximate solution to this equation by considering a non-zero contribution to the source term from only the matter overdensity, giving the well known solution $\delta_c \propto \eta^2$. In the case of the ASCDM model, the scalar field contributes to the second term on the left-hand side but does not contribute appreciably to the source term on the right-hand side compared to the contribution from matter. In particular, since $\Omega_\phi$ begins to decrease at the onset of matter-radiation equality (see Fig.~\ref{fig:delH}), there is a scale dependent suppression of the density contrast.\footnote{In fact, for a quintessence field that continues to track the background matter component during matter domination, the solution to Eq.\ref{eq:delta-m} is given by $\delta_{c} \propto \eta^{2+\frac{5}{2}\left(-1+\sqrt{1-\frac{24}{25} \Omega_{\phi}}\right)}$ as opposed to the usual $\delta_c^{\Lambda CDM} \propto \eta^2$ (See Section III in \cite{FerJoyce}).} Modes that enter before matter-radiation equality experience the maximal suppression while modes that enter after asymptotically approach the $\Lambda$CDM solution. This can be seen clearly in the matter power spectrum in Fig. \ref{fig:mpk}. This effect is responsible for power suppression and leads to a decreased inference of the power spectrum normalization, $\sigma_8$. Thus, if $\omega_m$ does not increase significantly, $S_8 = \sigma_8 (\Omega_m/0.3)^{0.5}$ is also suppressed. This effect of power suppression has been noted previously in the literature as well \cite{Jennings_2010, Skordis_2002,Jerome-Martin-P-Brax, FerJoyce}.
 
However, the CMB angular power spectrum exhibits a radiation driving envelope \cite{Hu-Sugiyama-Rad-Driving} which is sensitive to the matter-radiation ratio. In particular, since radiation causes the Newtonian potential $\psi$ to decay, acoustic oscillations that occur for modes entering the horizon before matter-radiation equality see a boost in amplitude since $\psi$ is timed to decay when photons in the fluid reach maximum compression. Since the quintessence field effectively contributes to the radiation budget during the radiation dominated epoch, there is a shift in the driving envelope, the implications of which we discuss more extensively in Section 5.1. The same physics is also relevant for an increased Early Integrated Sachs-Wolfe (ISW) effect which affects modes that cross the horizon close to recombination when the photon visibility sharply increases. The net result is that the increased value of $\omega_{\rm m}$ expected from these effects counteracts, to some extent, efforts to alleviate the $H_0$ and $S_8$ tensions.

\section{Elements of our analysis}
\label{sec:4}
Here we discuss the technical details and tools used to arrive at our results, which may be of particular interest to those interested in reproducing our results or comparing with models of a similar nature. We also extensively document our data below since many of the experiments have a series of data releases (DR), and comparing fits and parameter estimates across models using different data sets can lead to erroneous conclusions, as is extensively discussed in \cite{h0olympics}. 
\subsection{Data}
\begin{enumerate}
    \item \textbf{Baryonic Acoustic Oscillations (BAO):} We use the BOSS BAO DR12 ``consensus'' data set \cite{Alam-BOSS-BAO-DR12}. The BAO-only data, which is a subset of the ``consensus" dataset, relies on reconstructing the BAO signal which assumes a cosmological model, namely $\Lambda$CDM. However, as outlined in \cite{White-Sherwin-BAO-Reconstruction-Errors}, moderate differences in the distance scale between the fiducial and underlying cosmology lead to negligible changes in the monopole peak position. Thus, while we have not explicitly implemented the reconstruction using the ASCDM model, we think it is reasonable to use the data nonetheless, as is done in the $H_0$ tension literature.  
    
    \item \textbf{Cosmic Microwave Background (CMB):} We use the Planck 2018 data \cite{Planck-Collab-likelihoods, Planck-Collab-lensing}, including lensing, low-$\ell$ TT (\texttt{commander}), EE (\texttt{simall}) and high-$\ell$ temperature and polarization power spectra. For the high-$\ell$ spectra, we make use of the \texttt{plik-lite} code which differs from the full \texttt{plik} likelihood in the number of nuisance parameters.  
    
    \item \textbf{Supernovae (SNe):} We use the \texttt{Pantheon}  \cite{Pantheon} compilation of the redshift and magnitudes of 1048 Type-1a SNe in the range $0.01<z<2.3$. 
    
    \item \textbf{Cepheid Calibrated Supernovae:} We use the SH$_0$ES \cite{Riess19} value of $H_0 = 74.03\pm1.42 \textnormal{ km/s/Mpc}$, which effectively acts as a prior on $H_0$. Note that  while this work was in its final stages of completion, the SH$_0$ES team announced new results \cite{Riess:2021jrx} where they report $H_0 = 73.04 \pm 1.04$ which further exacerbates the tension with the Planck-calibrated \lcdm\ value of $H_0$. While using this new result would slightly alter the numerical values for our parameter inferences,  it would not induce any qualitative or drastic quantitative change to the analysis that we present for the ASCDM scenario. Whenever we quantify a difference with ``SH$_0$ES'' we implicitly refer to the 2019 result \cite{Riess19} in particular.
    
    \item \textbf{Sunyaez-Zeldovich (SZ) Cluster Count}: We use the Planck SZ value of $\sigma_8(\Omega_m/0.27)^{0.3} = 0.782 \pm 0.010$ \cite{Planck-SZ} which is effectively a prior on $S_8$. 
    
    \item \textbf{Large-Scale Structure (LSS):} We use a joint constraint on $S_8$ as found by \cite{Joudaki-S8-prior} using data from the Dark Energy Survey (DES), KiDS, and VIKING-450 (along with cosmological priors from \texttt{DES-Y1} \cite{Troxel-18-S8-prior}). This joint constraint is in mild tension, at an $\approx3\sigma$ level, with the Planck+$\Lambda$CDM inference. Using only the $S_8$ constraint, as opposed to the full `$3\times 2$ point' analysis, was done for the purposes of computational simplicity. This was motivated by \cite{EDE-Hill-LSS}, who found that most of the constraining power from DES can be summarized into a Gaussian prior on $S_8$ for their Early Dark Energy (EDE) scenario, although we have not explicitly tested the validity of doing so for the ASCDM model. In any case, the statistical power of the constraint is small compared to the aforementioned datasets and, by itself, is not decisive in setting parameter inferences.
    
\end{enumerate}

\subsection{Parameter space, priors}
In order to obtain statistical constraints on the parameter space given the data, we vary the usual six cosmological parameters and additionally vary the quintessence parameter $\lambda$ (in Eq.~\ref{eq:AS}) so that the complete set of varied parameters is $\{\omega_b, \omega_{cdm}, \theta_s^*, n_s, A_s, \tau_{reio}, \lambda \}$. One might wonder whether one degree of freedom is able to capture the complicated dynamics of the scalar field; this is indeed the case since:
\begin{enumerate}
    \item The quintessence parameter $A$ primarily plays the role of establishing a minimum. In particular, for a given $\lambda$, a local minimum exists if and only if $A < 1/\lambda^2$ and $A \neq 0$ in order to have a false vacuum. We fix $A=0.0025$. Note that the choice of $A$ has a mild impact on the curvature of the minimum and therefore affects the amplitude of the late-time oscillations when the scalar field is settling into the minimum. We have checked, by fixing $A$ to various values, that the parameter inferences are not sensitive to (small) changes in $A$. However, it is possible that the late-time oscillations may ultimately serve as probes for testing models of scalar field potentials with local minima.
    \item For a given $\lambda$, the location of the minimum is primarily governed by $B$. But since, 
    \begin{equation}
    \rho_\phi^0 = 3H_0^2(1-\Omega_m^0) \approx V(\phi_{\rm min})
    \label{eqn:rhofi0}
    \end{equation}
     $\phi_{\rm min}$ (and therefore $B$) is set uniquely in terms of the other background parameters that are varied. Here a sub-script or super-script naught denotes the value of the parameter today and $\phi_{\rm min}$ is the location of the minimum of the potential in Eq.~\ref{eq:AS}.
    \item $V_0$ is passed as a shooting parameter to CLASS for any fine-tuning required to satisfy Eqn.~\ref{eqn:rhofi0}. In reality, the need for this parameter is purely computational since if we vary $H_0$ instead of $\theta_s^*$, we can invert Eqn.~\ref{eqn:rhofi0} to set the minimum at exactly the requisite value of $\phi$. However, since $H_0$ is itself a derived parameter and is not known \textit{a-priori} for each step in the Monte Carlo, one must use a fiducial value of $H_0$ in Eqn.~\ref{eqn:rhofi0} and then use $V_0$ as a tuning (or ``shooting") parameter.
    
    \item For the tracking solution that we are interested in, the scalar field equation of motion (Eq. \ref{eq:KG}) admits a large basin of attraction, such that the initial conditions of the field, $\phi_\textnormal{ini}$ and $\dot{\phi}_\textnormal{ini}$, do not need to be fine tuned. We set them deep in the radiation era such that $\rho_\phi(a_\textnormal{ini}) \lesssim \rho_{\textnormal{radiation} }(a_\textnormal{ini})$ and that $w_\phi(a_\textnormal{ini}) = 1/3$.
\end{enumerate}    
Finally, $\lambda$ is the only outstanding parameter in Eqn.~\ref{eq:AS} which we vary in the prior range $2\leq \lambda \leq 1/\sqrt{A} $. The lower bound comes from the consideration that a tracking/scaling solution in the radiation era exists only for $\lambda \geq 2$ \cite{FerJoyce}. The upper bound occurs for the reason outlined in item $1$ above. Note that larger values of $\lambda$ drive the minimum to lower values of $\phi$ which in turn increases the probability for the scalar field to tunnel through the local minimum, leading to ``cataclysmic'' Coleman-De Luccia bubbles as discussed originally in \cite{Coleman-DeLuccia} and specifically for a class of quintessence potentials in \cite{Bubbles-Jochen-Weller}.While intriguing, we have excluded this region of parameter space that would postulate a probable end to accelerated expansion soon after dark energy becomes dominant\footnote{Similarly, we also exclude the  behaviors considered in~\cite{10.1046/j.1365-8711.2000.03778.x} which end the acceleration via classical processes.}. For this reason, we set $A=0.0025$, so that $\lambda \leq 20$. This also has the technical advantage that it prevents the sampler from wandering into the high likelihood region that is degenerate (asymptotically) with a cosmological constant. We also note that the dynamical parameter of the theory ($\lambda$) is $\mathcal{O}(10)$ (in Planck units). Some have found this feature, combined with the large basin of attraction for the initial conditions, philosophically appealing (e.g. see Section II of \cite{Jerome-Martin-P-Brax}).

\subsection{Software}
We use the Cosmic Linear Anisotropy System Solver (CLASS) \cite{CLASS-paper} as our Einstein-Boltzmann solver to calculate all quantities of cosmological interest. For sampling the parameter space, we use MontePython \cite{MontePython1,MontePython2}. For exploring the ASCDM model, we use the MultiNest \cite{MultiNest1,MultiNest2,MultiNest3} sampler as implemented in MontePython via PyMultiNest \cite{PyMultiNest}. Finally, we use GetDist\cite{getdist} to generate all posterior distribution plots.  
\section{Results}
\label{sec:5}
\begin{figure*}
\includegraphics[width=16cm]{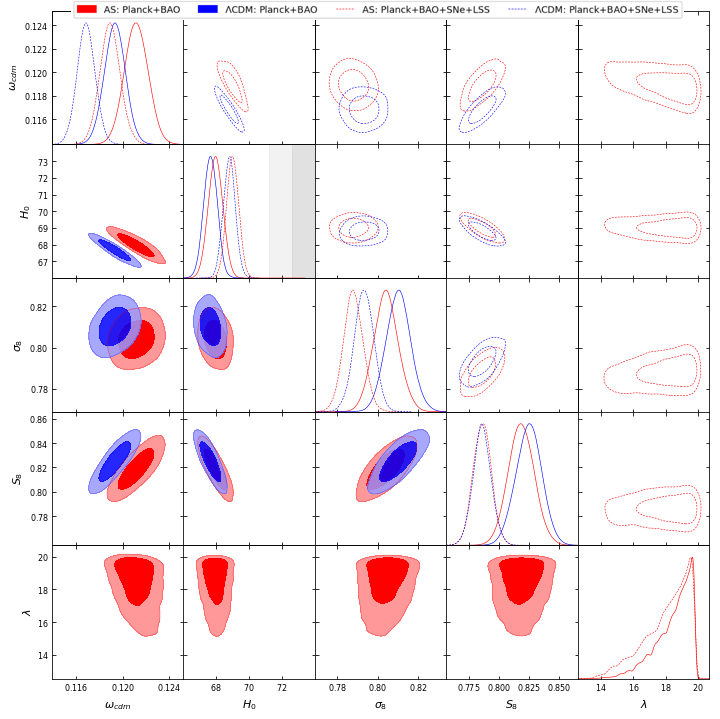}
\caption{Posteriors for the ASCDM and \lcdm\ models. Lower triangle (solid) shows the results using only Planck+BAO data, while posteriors in the upper triangle (dashed) also include LSS (both the $S_8$ prior from \cite{Joudaki-S8-prior} and Planck SZ data) and Pantheon (SNe) data. Here we show only the most relevant parameters for the cosmological tensions discussed in the text. The corresponding numerical values for the mean, best-fit, and credible intervals for the parameters can be found in Table~\ref{tab:parameter-inferences} while the $\chi^2$ values can be found in Table~\ref{tab:chi2-table}.
\label{fig:AS-LCDM-pb-mix}}
\end{figure*}

\begin{figure*}
\includegraphics[width=16cm]{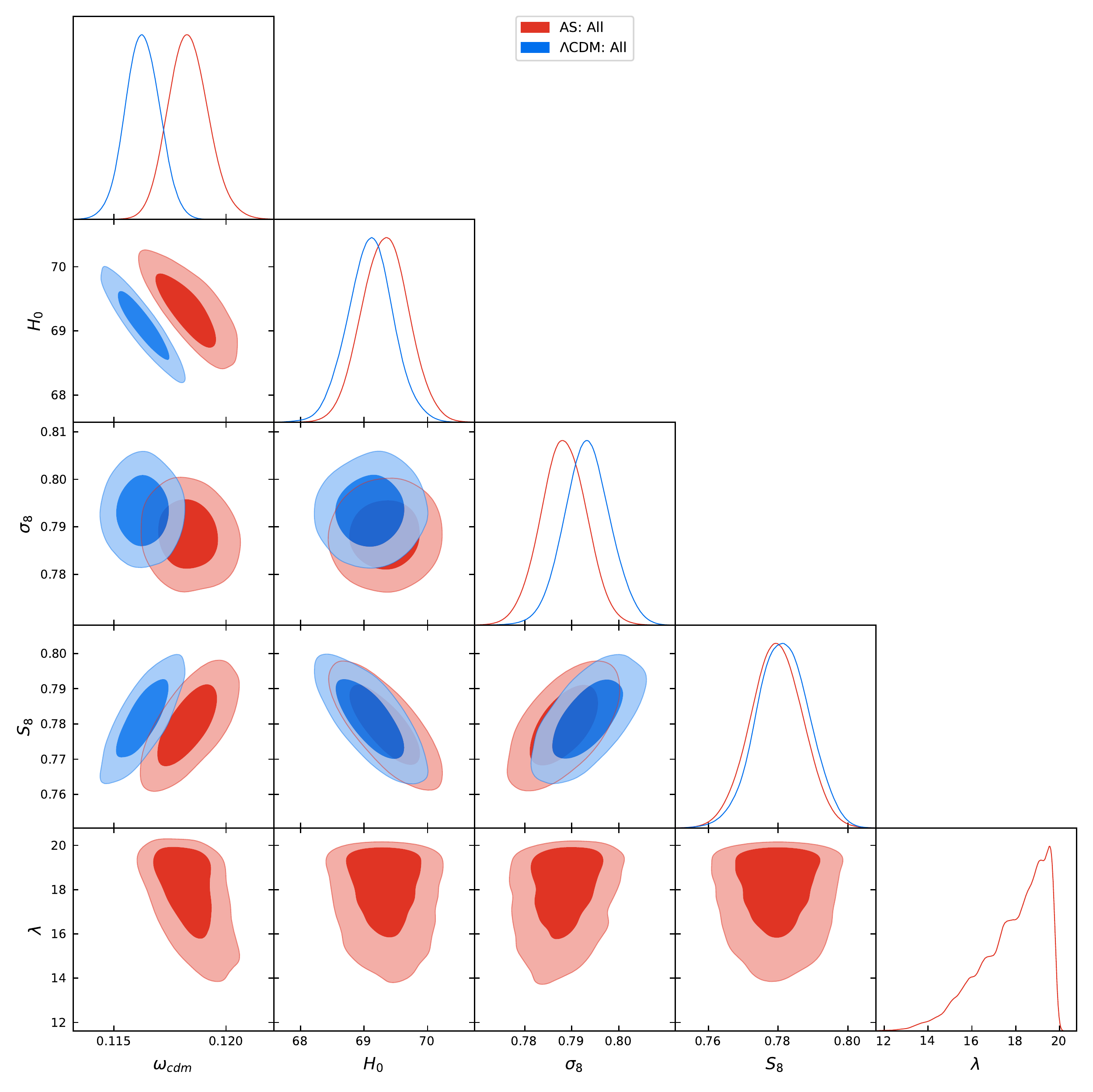}
\caption{Posteriors for the ASCDM and \lcdm\ models when including all data (discussed in Sect.~\ref{sec:4}), including SH$_0$ES. Here we show only the most relevant parameters for the cosmological tensions discussed in the text. When using this more holistic set of data, we find $\Delta \chi^2 \approx +1$.  The corresponding numerical values for the mean, best-fit, and credible intervals for the parameters can be found in Table~\ref{tab:parameter-inferences} while the $\chi^2$ values can be found in Table~\ref{tab:chi2-table}.
\label{fig:AS-LCDM-all}}
\end{figure*}

In Fig.~\ref{fig:AS-LCDM-pb-mix}-\ref{fig:AS-LCDM-all} we show constraints on the 6+1 parameter space of our model for various combinations of the data. Perhaps the first thing to notice is that these data do not prefer the ASCDM model. The posterior probability density peaks right at our prior cutoff at $\lambda=20$ for all the data combinations we examine (in the limit of large $\lambda$ the quintessence field contribution becomes more and more like that of a cosmological constant). For Planck+BAO, the quality of the best fit to the data is considerably worsened with $\Delta \chi^2 \approx +6$. When considering all datasets (Section 3.1.1), we find that the fit quality is comparable to that of $\Lambda$CDM with $\Delta \chi^2 \approx +1$. The fit to the various data are summarized in Table \ref{tab:chi2-table} and the numerical values for the best-fit parameters, along with the corresponding mean and credible intervals, can be found in Table \ref{tab:parameter-inferences}.

Although not preferred, by opening up this extension to finite $\lambda$, there is both a shift upward of approximately $1\sigma$ in the posterior distribution of $H_0$ and a similar downward shift in the posterior distribution of $\sigma_8$ compared to $\Lambda$CDM, alleviating both tensions reviewed in Sect.~\ref{sec:2}. We now turn to the question of why we do not get bigger shifts than this, as we had hoped we would, and why the constraints on $\lambda$ are as tight as they are; i.e., why do the data disfavor lower values of $\lambda$?

To answer these questions we have found it useful to compare to what happens in a much more thoroughly-studied $\lcdm$ extension, the one-parameter extension with a variable effective number of neutrino species, $N_{\rm eff}$ (\cite{Hu:1996,bashinsky04,hou2013,cyr-racine2022} and Ge et al. in preparation). So we begin our answer to these questions by reviewing this case of \lcdm + $N_{\rm eff}$. 

In the \lcdm\ + $N_{\rm eff}$ model it is possible to vary a subset of the 7 parameters of the model so that there is a uniform scaling of $H(z) \rightarrow \alpha H(z)$~\cite{cyr-racine2022}. This scaling transformation is of interest to us since both CMB temperature and polarization anisotropies and BAO observables are approximately invariant under it. The transformation approximately preserves two very important angular scales: the angular extent today of the sound horizon on the last scattering surface, $\theta_{\rm s}$, and the angular extent of the comoving size of the horizon at matter-radiation equality, projected from the last-scattering surface, $\theta_{\rm EQ}$. The CMB power spectra are very sensitive to these two scales because the latter largely determines the separation between peaks and the former determines the angular-scale-dependent impact of ``radiation driving''  \cite{hu2001}.

Radiation driving \cite{Hu-Sugiyama-Rad-Driving} is a dramatic effect, whereby potential decay at horizon crossing greatly increases the oscillation amplitude of the fluid for modes that enter deep in the radiation-dominated era. The influence of non-relativistic dark matter is to preserve the potential and thus for modes that cross when the fraction of dark matter density is higher the amplitude increase is reduced.  The dependence of this amplitude boost on angular scale is referred to as the radiation-driving ``envelope.''  

The symmetry under scaling of $H(z)$ is only approximate due to a number of effects, photon diffusion perhaps being chief among them. For a full explanation see \cite{cyr-racine2022} and Ge et al. (in preparation). But the ability to make adjustments that preserve $\theta_s$ and $\theta_{\rm EQ}$ lead to constraints on $N_{\rm eff}$ being much looser than they otherwise would be. This is an important point, as this scaling transformation $H(z) \rightarrow \alpha H(z)$ is not achievable in the ASCDM model and this plays a role in our understanding of the strong constraints we have found on the model.

\begin{figure}
    \centering
      \subfigure{\includegraphics[clip,width=3in]{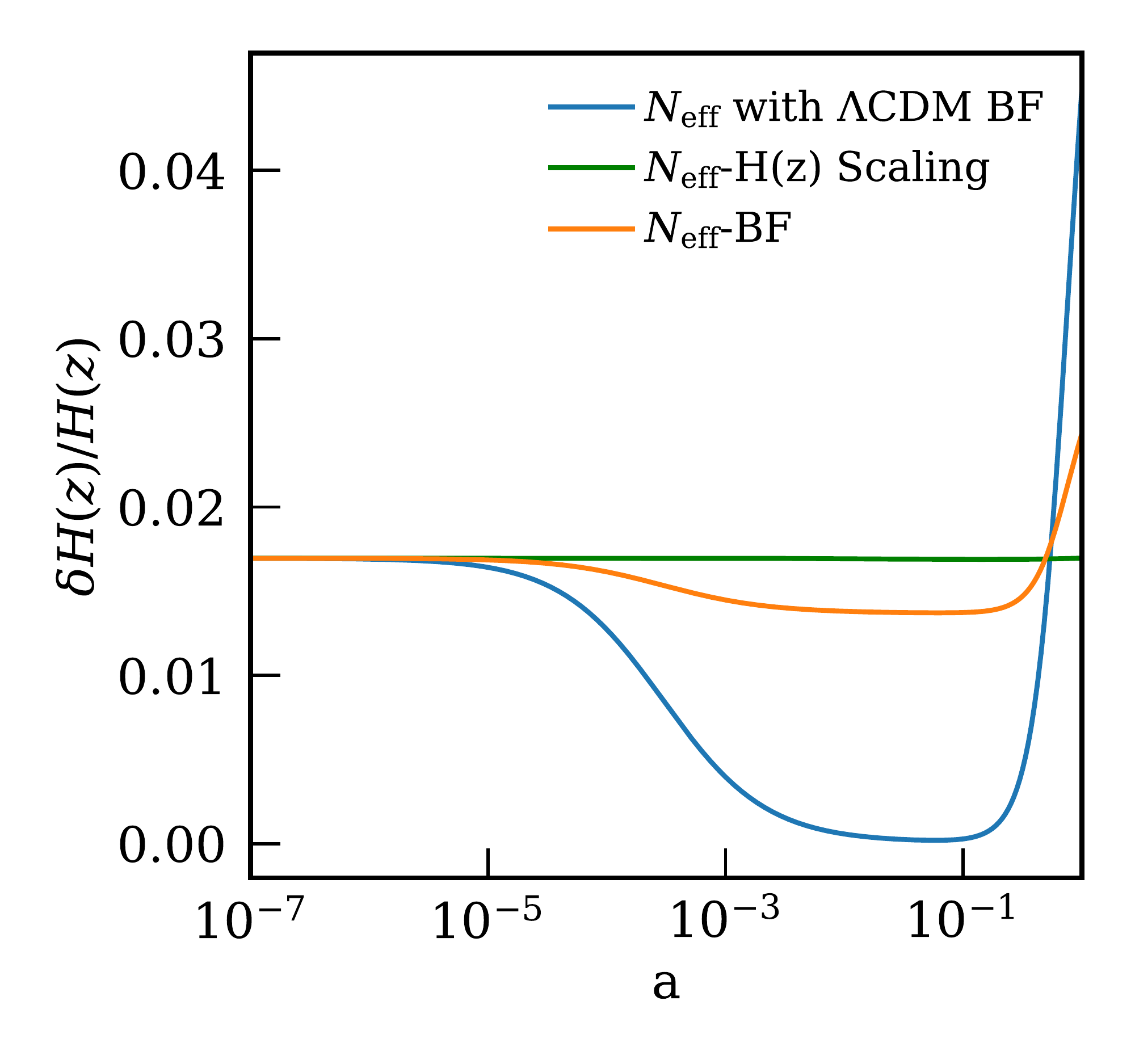}}
      \vspace{-1.5\baselineskip}
      
    \subfigure{\includegraphics[clip,width=3in]{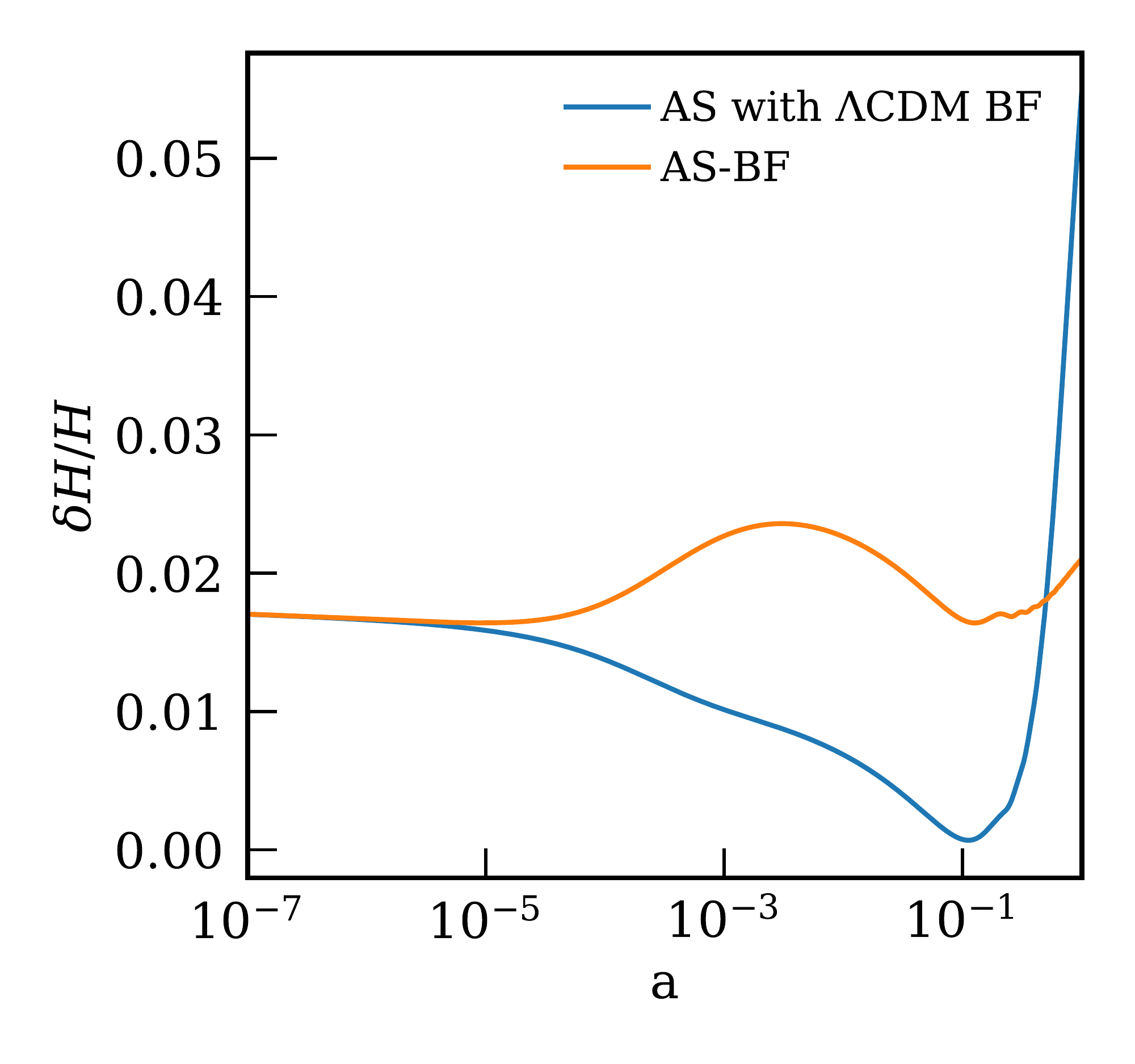}}
    \vspace{-1.5\baselineskip}
    \subfigure{\includegraphics[clip,width=3in]{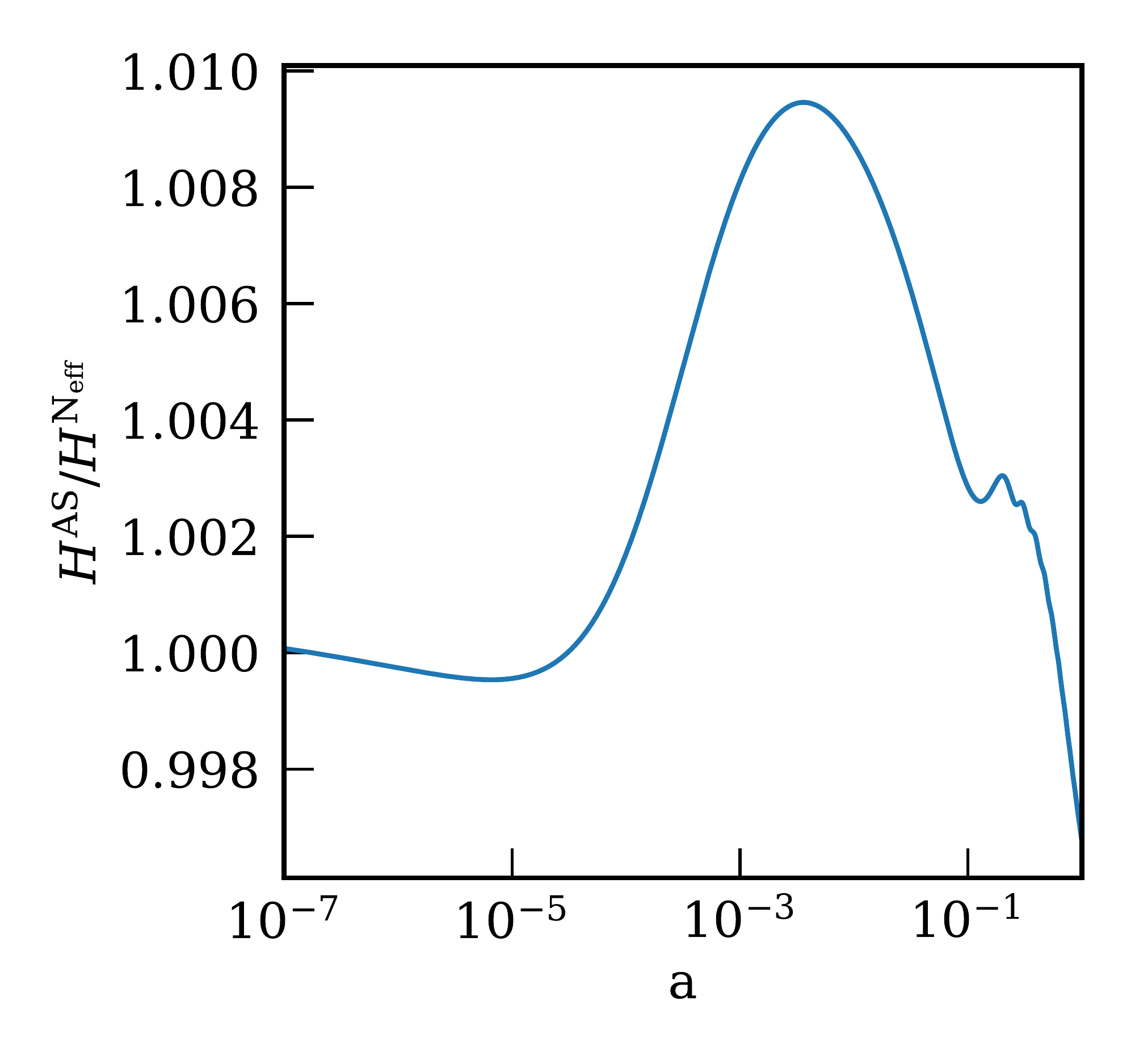}}
    
    \caption{For the cases discussed in Section 5, we show the change in expansion rate, relative to the best-fit \lcdm\ model, for the \lcdm\ + $N_{\rm eff}$ model with $N_{\rm eff}=3.3$ (top panel) and the ASCDM model with $\lambda=10.5$ (middle panel). The bottom panel shows the relative expansion rate between the best-fit AS and \lcdm\ + $N_{\rm eff}$ models. }
    \label{fig:As-Neff-scaling}
\end{figure}

\begin{figure}
\includegraphics[width=9cm]{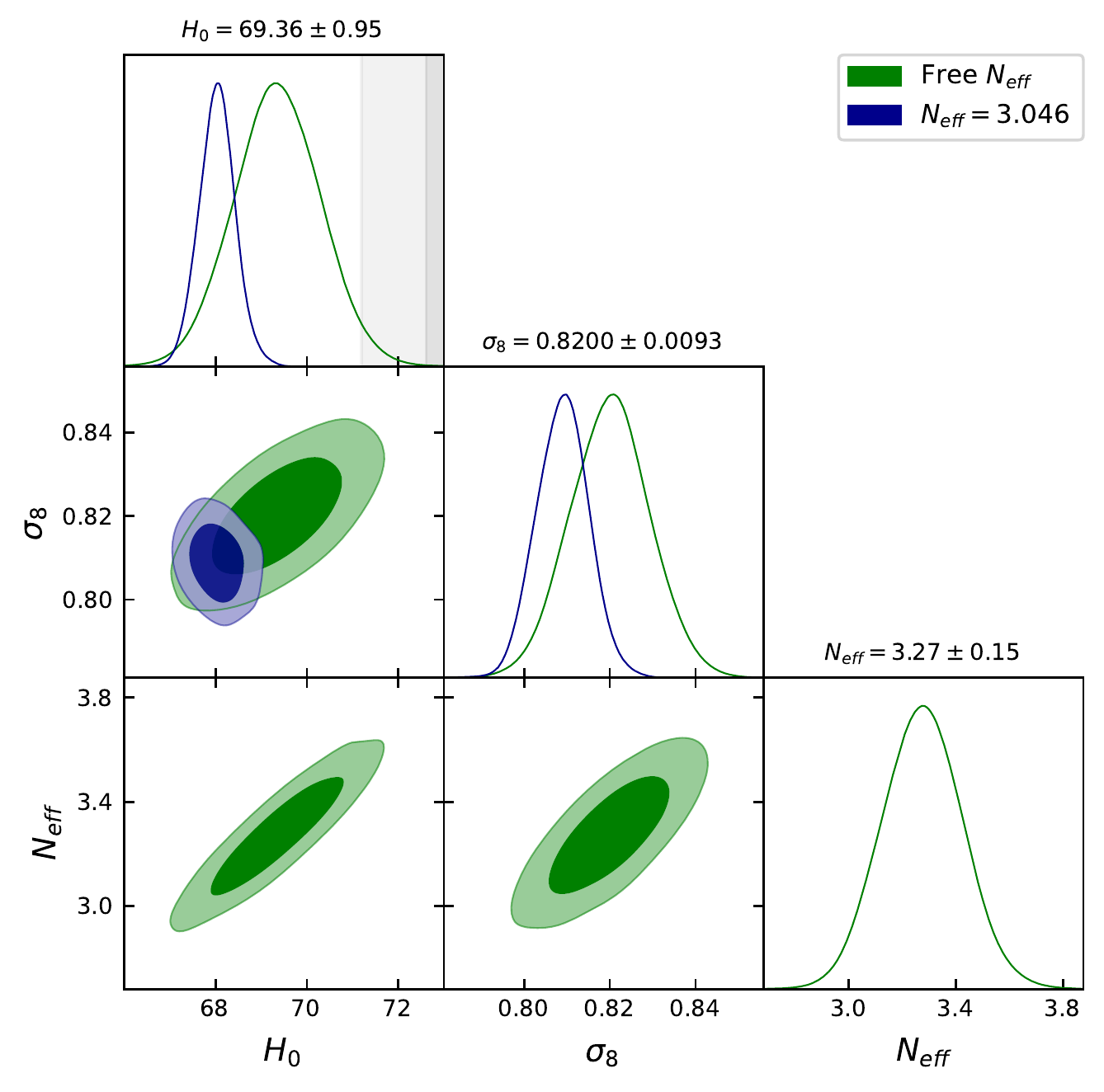}
\caption{Posterior distributions for a subset of the \lcdm\ + $N_{\rm eff}$ model using publicly available chains from the Planck Legacy Archive \cite{PLA} which uses Planck, BAO, Riess18. Evidently, allowing $N_{\rm eff}$ to vary freely in the presence of a $H_0$ prior from Riess18~\cite{Riess18} leads to increased $H_0$, but at the cost of increased $\sigma_8$.}
\label{fig:Neff}
\end{figure}

Before fully turning to the ASCDM model, let's look at how $H(z)$ changes in the \lcdm\ + $N_{\rm eff}$ model with increased $N_{\rm eff}$. 
In Fig.~\ref{fig:As-Neff-scaling} (top panel) we plot the fractional difference of $H(z)$ from the best-fit \lcdm\ model (given Planck + BAO data) for three cases, all with $N_{\rm eff}$ fixed to 3.3. The first case has all the other parameters ($\tau$,\ $n_s$,\ $A_s$,\ $\omega_b$,\ $\omega_c$, and $\theta_{\rm s}$) fixed to their best-fit \lcdm\ values, the second case is the same but with $\omega_c$ and $\omega_\Lambda$ increased in just the amount needed to deliver $H(z) \rightarrow \alpha H(z)$, and the third case is the best-fit (given Planck + BAO and the constraint that $N_{\rm eff}=3.3$). We choose this particular value of $N_{\rm eff}$ because it is approximately the best-fit value that $N_{\rm eff}$ is naturally driven towards if one includes the SH$_0$ES prior on $H_0$ (along with Planck and BAO data) in the \lcdm\ + $N_{\rm eff}$ (with $N_{\rm eff}$ allowed to vary), as shown in Fig.~\ref{fig:Neff}.

The blue curve ($N_{\bf eff} = 3.3$, other parameters at best fit $\Lambda$CDM  values) illustrates the apparent promise of this modification, in that $H_0$ in that case is larger than for \lcdm. Naturally, this case gives an overall much poorer fit than \lcdm.    
We see that the best-fit $\delta H(z)/H(z)$ curve (orange) is indeed quite close to that of the scaling model (green). The symmetry-breaking effects have led to a mild departure from scaling for the best fit. Because the preference is for an $\omega_{\rm m}$ increase that is a bit smaller than the scaling value, to preserve $\theta_s$, $\omega_\Lambda$ has to increase beyond the scaling value, boosting $H_0$ above the scaling value. The symmetry breaking effects (and the consistency of the data with \lcdm ) also lead to a fit quality that is worse than the \lcdm\ case. We find $\Delta \chi^2 \approx 3$. The increase in $H_0$ is considerably lower than the promise offered by the blue curve. 

Now let's look at the same scenario with the ASCDM model in Fig.~\ref{fig:As-Neff-scaling} (middle panel). To compare with $N_{\rm eff} = 3.3$ we choose $\lambda = 10.5$ because it gives a similar contribution to the expansion rate deep in the radiation-dominated era. For the ASCDM model there is no way to adjust its parameters to achieve $H(z)\rightarrow \alpha H(z)$, so we do not show a scaling case in this figure. We do show the other two cases studied for \lcdm\ + $N_{\rm eff}$: the one with $\omega_m$ and $\theta_{\rm s}$ fixed and the one with the parameters at their best-fit values (given Planck + BAO data and the constraint that $\lambda = 10.5$). 

The best-fit value of $\omega_m$ has increased over the \lcdm\  best-fit value by a similar amount in both our AS case and \lcdm\ + $N_{\rm eff}$ cases: by about 3\%. Such an increase is presumably driven by a need to preserve, as well as possible, the radiation-driving envelope. Like radiation, the AS field is a smooth component that contributes to the expansion rate and not to gravitational instability, thus contributing to potential decay. However, unlike with \lcdm\ + $N_{\rm eff}$, there is no single adjustment of $\omega_{\rm m}$ that can preserve $\rho_{\rm m}/\rho_{\rm smooth}$ at all values of the scale factor, and thus the shape of the radiation-driving envelope is necessarily changed. 
Exacerbating this problem, the increase to $\omega_{\rm m}$ pushes back the epoch of matter-radiation equality, which makes the transition to smaller values of $w_\phi$ start at higher redshift, thereby boosting $\rho_\phi$ at $z > z_{\rm EQ}$ and prior to dark-energy domination. This increase to $\rho_\phi$ raises $\delta H(z)/H(z)$ in the matter-dominated era, requiring the late-time dark energy density to come down so much that the boost to $H_0$ almost entirely disappears. This increase in $H(z)$ from a smooth component in the matter-dominated era is precisely what is responsible for the decrease in $\sigma_8$, as we discussed in Sect.~\ref{sec:3}. 

Even though the gain in $H_0$ is similar to the $N_{\rm eff}=3.3$ best-fit case, the quality of the fit is significantly worse: we find $\Delta \chi^2 \approx 30$, driven almost entirely by the Planck high-$\ell$ data. We suspect the origins of this degradation are in the change to the shape of the radiation-driving envelope. This change flows from the changed shape of $H(z)$, which is most clearly seen in the bottom panel of Fig.~\ref{fig:As-Neff-scaling} where we directly compare the AS best-fit $H(z)$ with the \lcdm\ + $N_{\rm eff}$ best-fit $H(z)$.  After $a \simeq 2 \times 10^{-5}$  there begins a rise in $H^{\rm AS}/H^{N_{\rm eff} }$, peaking with a nearly 1\% excess just after recombination. Recall that modes that cross the horizon ($k/a = H$) at matter-radiation equality project from the last-scattering surface into about $\ell = 140$. The corresponding $\ell$ values for modes that cross earlier scale nearly linearly with $1/a$, and thus the modes crossing when $a=2\times 10^{-5}$ project into about $\ell = 2000$. Modes crossing during this rise thus project into $\ell = 2000$ and lower, where the {\it Planck} data are very sensitive.


We see here an example of there being a whole function to match related to the transition from radiation domination to matter domination, rather than a single scale. The sensitivity of power spectra to this transition has been noted before; for example in \cite{hubblehunters} the authors noted that there are residuals to the best-fit \lcdm\ model, leading to slightly different $\omega_{\rm m}$ inferences depending on the $\ell$ range of data used, that could potentially be explained by departures from \lcdm\ in the decade or two prior to recombination. Apparently what we are seeing from the ASCDM model is differences with respect to \lcdm\ spectra, but not ones that look like the {\it Planck} residuals.  

Next, we turn to the importance of the damping scale, characterized by the angular scale $\theta_D$, in determining the parameter inferences for the ASCDM model. Here too we take inspiration from the \lcdm\ + $N_{\rm eff}$ model where allowing the primordial abundance of Helium, $Y_P$, to vary leads to considerable freedom in $\Delta N_{\rm eff}$ \cite{PCP18, Knox-neutrinos-damping-tail, cyr-racine2022, trouble-h0}. However, while this path seems appealing, the change to the damping scale seems unlikely to be the limiting factor in driving the posterior distribution of $\lambda$ towards the prior edge. We infer this by varying $Y_p$ (as opposed to calculating it using standard BBN)\footnote{ Since $\theta_D$/$\theta_s$ = $r_D$/$r_s$ $\propto H^{0.5}$, fixing $\theta_s$ will in general change $\theta_D$. Instead, one can get an independent handle on $\theta_D$ by varying $Y_p$. For example, in order to fix the angular scale $\theta_D$ while varying the expansion rate at recombination, one can set $Y_p$ using the prescription in e.g. \cite{Knox-neutrinos-damping-tail}\cite{cyr-racine2022}} in the ASCDM model and find that the parameter inferences and $\chi^2$ fit remain approximately unchanged, in contrast to the \lcdm\ + $N_{\rm eff}$ case.

\subsection{Comparison with Other Models}
From glancing at the posteriors in Fig.~\ref{fig:Neff}, one may wonder if the \lcdm+$N_{\rm eff}$ model (with $N_{\rm eff}$ allowed to vary) provides a solution to the $H_0$ tension. After all, that result differs from the SH$_0$ES result by only $\approx 2.7\sigma$.\footnote{Although, it should be pointed out that the chains used to produce the posteriors in Fig.~\ref{fig:Neff} use an older version of the SH$_0$ES result from \cite{Riess18}, as opposed to the one used in our results  \cite{Riess19}.} However, this is not a satisfactory solution for at least two reasons. First, the increased inference of $N_{\rm eff}$ (and therefore of $H_0$) occurs only due to the inclusion of the SH$_0$ES prior; if that is excluded, Ref.~\cite{PCP18} find $N_{\rm eff}=2.99^{+0.34}_{-0.33}$ and $H_0=67.3\pm1.1 {\rm km s^{-1} Mpc^{-1}}$, discrepant with SH$_0$ES at $\approx3.7\sigma$. Second, and more importantly for the comparison with the ASCDM model, the inference for $\sigma_8$ has increased so that the tension with LSS data is worsened, as explained in \cite{PCP15} (Section 6.4.2). 

This issue of an increased $\sigma_8$ accompanying an increased $H_0$ is not unique to the \lcdm+$N_{\rm eff}$ model and plagues several early-time proposals to remedy the $H_0$ tension \cite{Clark:2021hlo}. For example, there is the indirect effect of an increased $\sigma_8$ in the EDE model which, from the consideration of increasing $H_0$ ``efficiently'', does very well since the proposed axion-like field quickly decays around the time of recombination. However, for the case of EDE, \cite{EDE-Hill-LSS, EDE-Ivanov-LSS} point out that including the LSS data erases any preference for the model over $\Lambda$CDM (note, however, that the effect of LSS data is also obfuscated by prior volume effects as pointed out in \cite{EDE-prior-volume, EDE-laura-likelihood}).

Finally, typical ``tracking'' models of quintessence are also unable to resolve the $H_0$ tension, primarily because they track the dominant background component throughout radiation and matter dominated epochs. Thus, even though they are able to decrease $r_s^*$, the requisite change in $D_M^*$ is diluted throughout the post-recombination era as opposed to at $z\approx0$. One such classical Ratra-Peebles quintessence potential, $V(\phi) \propto \phi^{-\lambda}$ is explored by \cite{Ooba-Ratra-Sugiyama} and their results corroborate the analysis above. However, it is important to note that in the case of the Ratra-Peebles model, $\lambda>0$ leads to $w_\phi > -1$ so that a Monte Carlo parameter estimation that finds $\lambda$ consistent with $0$ might be probing the preference for a dark energy equation of state equal to (or less than) $-1$ today, instead of solely an aversion to tracking quintessence. For this reason, we repeat our analysis for the so-called ``Brane'' model of quintessence, first introduced in \cite{Skordis_2002}, which tracks the background component throughout the expansion history of the Universe and also has $w_\phi\approx-1$ today nearly independently of the value of $\lambda$. The results, elaborated on in Appendix A are consistent with the analysis above, indicating that such a solution is unable to increase $H_0$ beyond the $\Lambda$CDM value. In particular, since in the presence of a non-negligible quintessence component during radiation domination the Planck data drives $\omega_{\rm cdm}$ to a higher value (because of the radiation-driving envelope), such a model leads to a \textit{decreased} inference of $H_0$ compared to $\Lambda$CDM. 

This same effect is, presumably, also responsible for the decreased inference of $H_0$ in the ``assisted quintessence'' scenario studied by \cite{H0-Assistance-Caldwell-Sabla}. Though that is not a typical tracking model but instead has multiple fields that thaw from Hubble friction at different epochs, leading to contributions to $H(z)$ throughout the matter dominated epoch. While the authors find the exacerbation of the tension ``a surprise'', perhaps it is not so surprising in light of the analysis above that, if the scalar field perturbations are consistently accounted for, $H_0$ is suppressed. 

\section{Other Prospects for Quintessence}
\label{sec:6}
Even though the ASCDM model can not resolve the $H_0$ tension by itself, we use this as an opportunity to chart future paths which we could not explore here but which nonetheless are interesting avenues to consider in the search for a concordant model of cosmology.

So far, we have focused on a proposal to increase $H_0$ by means of an ``early-time solution'', namely one that reduces the size of the sound horizon at recombination. However, some have considered using late-time changes to the expansion rate as a possible remedy to the $H_0$ tension\footnote{Here we use the phrase ``$H_0$ tension'' loosely. Of course, for models that change $H(z)$ at low redshifts ($z\ll1$) one must consider the discrepancy in the absolute magnitude ($M_b$) calibration of SNe from the SH$_0$ES distance-ladder versus that derived from the distance-luminosity relationship of a cosmological model, as has been pointed out by several authors \cite{ToH0OrNotToH0, Camarena-21, HuBenevento:2020}. } (e.g. \cite{HuBenevento:2020,Alestas:2021, Weiqiang-Valentino:2020} none of which ameliorate the tension; and \cite{Alestas:2021luu,Nunes:2021zzi}). While we do not believe that it is possible for changes at late times to fully resolve the $H_0$ tension (due to constraints from BAO and Pantheon in particular), it remains a possibility that they could reduce the tension to some degree. This opens up the possibility of a combined early and late time solution (e.g. \cite{Alestas:2021,Sekiguchi:2020teg}; see also \cite{Vagnozzi:2021tjv} for arguments in favor of a combined solution). We note here that scalar field models offer the possibility of such a combined solution, since they can be dynamically important not only at early times (as in the case we study here), but also at late times (e.g. due to dark-sector coupling \cite{Nunes:2021zzi, Kumar:2016zpg, Lucca:2021eqy} ).

Another interesting scenario to consider is the coupled dark energy scenario studied by Gomez-Valent et. al. \cite{Gomez:Coupled-DE-H0}. They report an increased inference of $H_0$, despite using the Ratra-Peebles tracking quintessence potential which, at the background level, is  not a promising candidate (for the same reasons pointed in Section 5.1 and Appendix A) for alleviating the tension with SH$_0$ES. They also report an increased inference of $S_8$, which would likely lower the preference for the model when LSS data are also taken into account. It would be interesting to repeat their analysis using the ASCDM model which can provide an additional boost to $H_0$ at the background level and also mitigate the increase in $S_8$. Similarly, refs. \cite{Nunes:2021zzi, Valentino2021combined} also report that the $H_0$ tension is alleviated in the (phenomenological) dark sector interaction cosmology that they consider. 

Yet another avenue is to instead consider a ``k-essence'' field, which differs from the quintessence scenario via a non-canonical kinetic term. Crucially, the difference in the kinetic term allows the scalar field to have a variable sound speed\cite{essentials-of-k-essence-picon-mukhanov-steinhardt, DeDeo-Caldwell-Steinhardt-k-essence}. This \textit{may} allow a k-essence model to circumvent the constraint stemming from the radiation driving effect if the sound speed remains small during radiation domination. Conversely, it can also be seen as a test of the extent to which the fixed $c_s^2=1$ constraint is responsible for driving the posteriors in the quintessence scenario.

\section{Conclusions}
\label{sec:7}
A solution that can satisfactorily relieve both the $H_0$ and $\sigma_8$ tensions continues to elude  cosmologists. In the search for such a solution, we have found interesting ramifications of adding a quintessence component to the Universe. While other studies of quintessence in light of recent cosmological data have reached conclusions similar to ours \cite{Ooba-Ratra-Sugiyama, H0-Assistance-Caldwell-Sabla,impact-of-theoretical-priors}, our work offers several unique features that lead to new insights. Previous works typically consider models that scale with the dominant component during both the radiation and matter eras. We chose to focus on the ASCDM model in particular because it tracks the background component only during the radiation dominated epoch, allowing it to give a higher $-\delta r_s$ to $\delta H_0$ ratio than typical tracking models. In this sense, such a model provides, a-priori, the best-case early universe scenario for a single field quintessence field in ameliorating the $H_0$ tension. Furthermore, even though we varied just a single parameter, we saw how the AS quintessence exhibits behaviors that are not captured by simple phenomenological parameterizations of dark energy.  This illustrates the importance of not relying completely on phenomenological characterizations when analysing the observable effects of quintessence.   The ASCDM model also compares favorably to  others when it comes to the $\sigma_8$ tension, which is exacerbated in \lcdm\  + $N_{\rm eff}$ and most EDE models.

However, ultimately, the ASCDM proposal provides a poorer fit than $\Lambda$CDM, particularly to the Planck high-$\ell$ temperature and polarization data at the level of $\Delta \chi^2 \approx +4$, although the total fit to a more extensive set of data that includes LSS, SNe, and the SH$_0$ES measurements provides a fit comparable to $\Lambda$CDM at $\Delta \chi^2 \approx +1$. In doing so, the ASCDM proposal can lower $\sigma_8$ by $\approx 1-2 \sigma$ compared to $\Lambda$CDM, bringing concordance with the LSS measurements, while also increasing $H_0$ by $\approx 1 \sigma$, which nonetheless continues to be in tension with the SH$_0$ES measurement at $3.2-4 \sigma$ (see Table~\ref{tab:chi2-table}). 

There has been some emphasis in the literature on the importance of the three angular scales, $\theta_s$, $\theta_{\rm EQ}$, and $\theta_d$ and the need for alternative models to not have these scales depart by too much from their \lcdm\ best-fit values \cite{Hu:1996,hu2001,bashinsky04,hou2013,follin15,poulin2019,hubblehunters}. We have seen here an illustration that controlling these variables can be insufficient for maintaining a good fit to the Planck high-$\ell$ data. We have speculated that this is due to the fact that the physics behind the $\theta_{\rm EQ}$ requirement has to do with the radiation-driving envelope, which is in general a whole function not specified by a single number, and that the shape of this envelope in the ASCDM model does not lead to a good fit to the data, no matter the value of $\theta_{\rm EQ}$. This is in contrast to the \lcdm\ + $N_{\rm eff}$ model space where it is easier to maintain the shape of the radiation-driving envelope because the ratio of matter to radiation can be preserved at all values of redshift. 

We remind the reader that we use a very limited LSS dataset in the form of a Gaussian prior on the $S_8$ summary parameter. While we have not tested this ourselves, it may well be the case that including a more extensive set of data from various weak lensing, galaxy clustering, SZ catalogues etc. may tip the total fit in favor of the AS, or some other AS-like, potential. Indeed, there may already be hints of a worsening tension in $\sigma_8$, since the Hyper Suprime-Cam (HSC) results probe a different degeneracy direction in the ($\Omega_m$,$S_8$) plane than previous LSS probes \cite{HSC_2021}, and as these constraints improve it may prove beneficial to revisit this work. 

The observable impacts of AS quintessence are effectively controlled by one parameter, that is of $\mathcal{O}(10)$ (Planck units), and we regard that simplicity as a strength of the model.   Perhaps a bit hypocritically, but certainly following strong traditions in theoretical physics, we remain curious about the more complicated cases. Combining the early-time effects discussed in Sect.~\ref{sec:3} with the the methods we mention in Sect.~\ref{sec:6} for modifying the late-time dynamics could potentially provide a more satisfactory solution to the $H_0$/$\sigma_8$ tensions.

\begin{table*}[htb]
\begin{ruledtabular}
\begin{tabular}{p{2cm}|cc |cc |cc}
 &$\Lambda$CDM &AS &$\Lambda$CDM & AS &$\Lambda$CDM & AS \\
\hline
\texttt{High-$\ell$ TTTEEE\footnotemark[1]} & 584.23 & 588.75 & 594.00 &597.69 & 596.43 & 600.35\\
\texttt{Low-$\ell$ EE\footnotemark[1]}& 396.42 & 396.85 & 395.71 & 395.71 & 395.63 & 395.88\\
\texttt{Low-$\ell$ TT\footnotemark[1]} & 23 & 23.33 & 22.02 & 22.22 &21.66 & 21.97\\
Lensing\footnotemark[1] & 8.78 & 8.99 & 13.15 & 14.10 & 13.88 & 14.75\\
BOSS DR12 & 6.57 & 7.44 & 6.48 & 5.99 & 7.14 & 6.66\\
Pantheon & &  & 1026.77 & 1026.71 & 1026.77 & 1026.74\\
Planck SZ & &  & 11.73 & 10.20 & 9.38 & 7.68\\
DES-KV450  &  &  & 2.82 & 2.71 & 2.11 & 1.85\\
SH$_0$ES &  &  &  & & 12.31 & 11.03\\
\hline
Total $\chi^2$  & 1019.00 & 1025.36 & 2072.68 & 2075.33 & 2085.31 & 2086.91   \\
\hline
$H_0$ & $67.66\pm 0.41$ & $67.97\pm 0.45$ & $68.80\pm 0.37$ & $69.01\pm 0.37$ & $69.10\pm 0.36$ & $69.33\pm 0.37$\\
$\sigma_8$ & $0.8102\pm 0.0061$  &  $0.8041\pm 0.0061$ & $0.7933\pm 0.0049$ & $0.7880\pm 0.0049$& $0.7933\pm 0.0049$ & $0.7883\pm 0.0049$\\
\end{tabular}
\end{ruledtabular}
\footnotetext[1]{Planck 2018}
\caption{$\chi^2$ fit corresponding to the best-fit values (stated in Table~\ref{tab:parameter-inferences}) for the data we consider.}
\label{tab:chi2-table}

\end{table*}

\setlength{\tabcolsep}{0.5em} 
{\renewcommand{\arraystretch}{1.4}
\begin{table*}
    \centering
    \begin{tabular} { l |c|c|c}
    
     Parameter &  Planck + BAO & + SNe+LSS & + SH$_0$ES\\
    \hline
    $\omega_b     $& $ 2.247\pm 0.014 (2.244)$     &    $ 2.264\pm 0.013 (2.264)$    &    $ 2.271\pm 0.013 (2.270)$      \\ 
    $\omega_{cdm}     $& $ 0.1212\pm 0.0010 (0.1211)$    &    $ 0.11891\pm 0.00089 (0.1186)$    &    $ 0.11832^{+0.00083}_{-0.00093} (0.1178)$      \\ 
    $100*\theta_s     $& $ 1.04103\pm 0.00031 (1.04114)$    &    $ 1.04108^{+0.00035}_{-0.00031} (1.04120)$    &    $ 1.04114^{+0.00035}_{-0.00030} (1.04134)$      \\ 
    $\ln(10^{10}A_s)    $& $ 3.054^{+0.014}_{-0.015}(3.051)$    &    $ 3.032\pm 0.013(3.032)$    &    $ 3.038\pm 0.013(3.036)$      \\ 
    $n_s     $& $ 0.9665\pm 0.0037 (0.9661)$    &    $ 0.9705\pm 0.0036(0.9708)$    &    $ 0.9724\pm 0.0036(0.9729)$      \\ 
    $\tau_{reio}     $& $ 0.0586^{+0.0069}_{-0.0077}(0.0576)$    &    $ 0.0513\pm 0.0069(0.0520)$    &    $ 0.0545\pm 0.0068(0.0544)$      \\ 
    $\lambda     $& $ 18.5^{+1.4}_{-0.39}(19.8)$    &    $ 18.1^{+1.8}_{-0.55}(19.8)$    &    $ 17.9^{+1.9}_{-0.63}(19.8)$      \\ 
      & $\lambda>16.2$ & $\lambda>15.2$ & $\lambda>15.1      $\\
    \hline
    $H_0     $& $ 67.97\pm 0.45(67.83)$    &    $ 69.01\pm 0.37(68.92)$    &    $ 69.33\pm 0.37(69.31)$      \\ 
    $\Omega^0_\phi     $& $ 0.6875\pm 0.0060(0.6866)$    &    $ 0.7013\pm 0.0046(0.7013)$    &    $ 0.7052\pm 0.0045(0.706)$      \\ 
    $\sigma_8     $& $ 0.8041\pm 0.0061(0.805)$    &    $ 0.7880\pm 0.0049(0.7897)$    &    $ 0.7883\pm 0.0049(0.7894)$      \\ 
    $S_8     $& $ 0.819\pm 0.011(0.820)$    &    $ 0.7862\pm 0.0076(0.788)$    &    $ 0.7796\pm 0.0074(0.779)$      \\ 
    $M     $& &$ -19.381\pm 0.010 (-19.382)$    &    $ -19.372\pm 0.010 (-19.372)$  \\
    \hline
    \end{tabular}
    
    \caption{Parameter inferences for the three sets of data corresponding to the posteriors in Fig.~\ref{fig:AS-LCDM-pb-mix}-\ref{fig:AS-LCDM-all}. For each parameter we list the mean $\pm 68\%$ credible interval (bestfit). Since the posterior for $\lambda$ is highly non-Gaussian, we also state the $95\%$ highest posterior density interval.}
    \label{tab:parameter-inferences}
\end{table*}

\section*{Acknowledgments}
A. Adil would like to thank Anton Chudaykin, Fei Ge, Michael Meiers, and Constantinos Skordis for helpful discussions, as well as Tyler Erjavic for access to computational resources. 
A. Adil and L. Knox were partially supported by a grant from the Theory Frontier program of the DOE Office of Science.

\bibliography{this_paper}

\begin{thebibliography}{97}
\expandafter\ifx\csname natexlab\endcsname\relax\def\natexlab#1{#1}\fi
\expandafter\ifx\csname bibnamefont\endcsname\relax
  \def\bibnamefont#1{#1}\fi
\expandafter\ifx\csname bibfnamefont\endcsname\relax
  \def\bibfnamefont#1{#1}\fi
\expandafter\ifx\csname citenamefont\endcsname\relax
  \def\citenamefont#1{#1}\fi
\expandafter\ifx\csname url\endcsname\relax
  \def\url#1{\texttt{#1}}\fi
\expandafter\ifx\csname urlprefix\endcsname\relax\def\urlprefix{URL }\fi
\providecommand{\bibinfo}[2]{#2}
\providecommand{\eprint}[2][]{\url{#2}}

\bibitem[{\citenamefont{Albrecht and Skordis}(2000)}]{Albrecht:1999rm}
\bibinfo{author}{\bibfnamefont{A.}~\bibnamefont{Albrecht}} \bibnamefont{and}
  \bibinfo{author}{\bibfnamefont{C.}~\bibnamefont{Skordis}},
  \bibinfo{journal}{Physical Review Letters} \textbf{\bibinfo{volume}{84}},
  \bibinfo{pages}{2076} (\bibinfo{year}{2000}).

\bibitem[{\citenamefont{Skordis and Albrecht}(2002)}]{Skordis_2002}
\bibinfo{author}{\bibfnamefont{C.}~\bibnamefont{Skordis}} \bibnamefont{and}
  \bibinfo{author}{\bibfnamefont{A.}~\bibnamefont{Albrecht}},
  \bibinfo{journal}{Physical Review D} \textbf{\bibinfo{volume}{66}}
  (\bibinfo{year}{2002}), ISSN \bibinfo{issn}{1089-4918},
  \urlprefix\url{http://dx.doi.org/10.1103/PhysRevD.66.043523}.

\bibitem[{\citenamefont{Peirone et~al.}(2017)\citenamefont{Peirone, Martinelli,
  Raveri, and Silvestri}}]{impact-of-theoretical-priors}
\bibinfo{author}{\bibfnamefont{S.}~\bibnamefont{Peirone}},
  \bibinfo{author}{\bibfnamefont{M.}~\bibnamefont{Martinelli}},
  \bibinfo{author}{\bibfnamefont{M.}~\bibnamefont{Raveri}}, \bibnamefont{and}
  \bibinfo{author}{\bibfnamefont{A.}~\bibnamefont{Silvestri}},
  \bibinfo{journal}{Phys. Rev. D} \textbf{\bibinfo{volume}{96}},
  \bibinfo{pages}{063524} (\bibinfo{year}{2017}), \eprint{1702.06526}.

\bibitem[{\citenamefont{Chevallier and Polarski}(2001)}]{CPL-1}
\bibinfo{author}{\bibfnamefont{M.}~\bibnamefont{Chevallier}} \bibnamefont{and}
  \bibinfo{author}{\bibfnamefont{D.}~\bibnamefont{Polarski}},
  \bibinfo{journal}{Int. J. Mod. Phys. D} \textbf{\bibinfo{volume}{10}},
  \bibinfo{pages}{213} (\bibinfo{year}{2001}), \eprint{gr-qc/0009008}.

\bibitem[{\citenamefont{Linder}(2003)}]{CPL-2}
\bibinfo{author}{\bibfnamefont{E.~V.} \bibnamefont{Linder}},
  \bibinfo{journal}{Phys. Rev. Lett.} \textbf{\bibinfo{volume}{90}},
  \bibinfo{pages}{091301} (\bibinfo{year}{2003}),
  \urlprefix\url{https://link.aps.org/doi/10.1103/PhysRevLett.90.091301}.

\bibitem[{\citenamefont{Ooba et~al.}(2019)\citenamefont{Ooba, Ratra, and
  Sugiyama}}]{Ooba-Ratra-Sugiyama}
\bibinfo{author}{\bibfnamefont{J.}~\bibnamefont{Ooba}},
  \bibinfo{author}{\bibfnamefont{B.}~\bibnamefont{Ratra}}, \bibnamefont{and}
  \bibinfo{author}{\bibfnamefont{N.}~\bibnamefont{Sugiyama}},
  \bibinfo{journal}{Astrophysics and Space Science}
  \textbf{\bibinfo{volume}{364}} (\bibinfo{year}{2019}), ISSN
  \bibinfo{issn}{1572-946X},
  \urlprefix\url{http://dx.doi.org/10.1007/s10509-019-3663-4}.

\bibitem[{DIV(2019)}]{DIVALENTINO2019100385}
\bibinfo{journal}{Physics of the Dark Universe} \textbf{\bibinfo{volume}{26}},
  \bibinfo{pages}{100385} (\bibinfo{year}{2019}), ISSN
  \bibinfo{issn}{2212-6864},
  \urlprefix\url{https://www.sciencedirect.com/science/article/pii/S2212686419302134}.

\bibitem[{\citenamefont{Sabla and
  Caldwell}(2021)}]{H0-Assistance-Caldwell-Sabla}
\bibinfo{author}{\bibfnamefont{V.~I.} \bibnamefont{Sabla}} \bibnamefont{and}
  \bibinfo{author}{\bibfnamefont{R.~R.} \bibnamefont{Caldwell}},
  \bibinfo{journal}{Phys. Rev. D} \textbf{\bibinfo{volume}{103}},
  \bibinfo{pages}{103506} (\bibinfo{year}{2021}), \eprint{2103.04999}.

\bibitem[{\citenamefont{Linden and Virey}(2008)}]{Linden-CPL-mapping}
\bibinfo{author}{\bibfnamefont{S.}~\bibnamefont{Linden}} \bibnamefont{and}
  \bibinfo{author}{\bibfnamefont{J.-M.} \bibnamefont{Virey}},
  \bibinfo{journal}{Phys. Rev. D} \textbf{\bibinfo{volume}{78}},
  \bibinfo{pages}{023526} (\bibinfo{year}{2008}), \eprint{0804.0389}.

\bibitem[{\citenamefont{Scherrer}(2015)}]{Scherrer-CPL-mapping}
\bibinfo{author}{\bibfnamefont{R.~J.} \bibnamefont{Scherrer}},
  \bibinfo{journal}{Phys. Rev. D} \textbf{\bibinfo{volume}{92}},
  \bibinfo{pages}{043001} (\bibinfo{year}{2015}), \eprint{1505.05781}.

\bibitem[{\citenamefont{Ratra and Peebles}(1988)}]{RP1988}
\bibinfo{author}{\bibfnamefont{B.}~\bibnamefont{Ratra}} \bibnamefont{and}
  \bibinfo{author}{\bibfnamefont{P.~J.~E.} \bibnamefont{Peebles}},
  \bibinfo{journal}{Phys. Rev. D} \textbf{\bibinfo{volume}{37}},
  \bibinfo{pages}{3406} (\bibinfo{year}{1988}),
  \urlprefix\url{https://link.aps.org/doi/10.1103/PhysRevD.37.3406}.

\bibitem[{\citenamefont{{Dvorkin} et~al.}(2022)\citenamefont{{Dvorkin},
  {Meyers}, {Adshead}, {Amin}, {Arg{\"u}elles}, {Brinckmann}, {Castorina},
  {Cohen}, {Craig}, {Curtin} et~al.}}]{dvorkin:2022}
\bibinfo{author}{\bibfnamefont{C.}~\bibnamefont{{Dvorkin}}},
  \bibinfo{author}{\bibfnamefont{J.}~\bibnamefont{{Meyers}}},
  \bibinfo{author}{\bibfnamefont{P.}~\bibnamefont{{Adshead}}},
  \bibinfo{author}{\bibfnamefont{M.}~\bibnamefont{{Amin}}},
  \bibinfo{author}{\bibfnamefont{C.~A.} \bibnamefont{{Arg{\"u}elles}}},
  \bibinfo{author}{\bibfnamefont{T.}~\bibnamefont{{Brinckmann}}},
  \bibinfo{author}{\bibfnamefont{E.}~\bibnamefont{{Castorina}}},
  \bibinfo{author}{\bibfnamefont{T.}~\bibnamefont{{Cohen}}},
  \bibinfo{author}{\bibfnamefont{N.}~\bibnamefont{{Craig}}},
  \bibinfo{author}{\bibfnamefont{D.}~\bibnamefont{{Curtin}}},
  \bibnamefont{et~al.}, \bibinfo{journal}{arXiv e-prints}
  \bibinfo{eid}{arXiv:2203.07943} (\bibinfo{year}{2022}), \eprint{2203.07943}.

\bibitem[{\citenamefont{Hu and White}(2009)}]{Hu-White-CMB-Damping-Tail}
\bibinfo{author}{\bibfnamefont{W.}~\bibnamefont{Hu}} \bibnamefont{and}
  \bibinfo{author}{\bibfnamefont{M.}~\bibnamefont{White}},
  \bibinfo{journal}{The Astrophysical Journal} \textbf{\bibinfo{volume}{479}},
  \bibinfo{pages}{568} (\bibinfo{year}{2009}).

\bibitem[{\citenamefont{{Hou} et~al.}(2013)\citenamefont{{Hou}, {Keisler},
  {Knox}, {Millea}, and {Reichardt}}}]{hou2013}
\bibinfo{author}{\bibfnamefont{Z.}~\bibnamefont{{Hou}}},
  \bibinfo{author}{\bibfnamefont{R.}~\bibnamefont{{Keisler}}},
  \bibinfo{author}{\bibfnamefont{L.}~\bibnamefont{{Knox}}},
  \bibinfo{author}{\bibfnamefont{M.}~\bibnamefont{{Millea}}}, \bibnamefont{and}
  \bibinfo{author}{\bibfnamefont{C.}~\bibnamefont{{Reichardt}}},
  \bibinfo{journal}{\prd} \textbf{\bibinfo{volume}{87}}, \bibinfo{eid}{083008}
  (\bibinfo{year}{2013}), \eprint{1104.2333}.

\bibitem[{\citenamefont{{Follin} et~al.}(2015)\citenamefont{{Follin}, {Knox},
  {Millea}, and {Pan}}}]{follin15}
\bibinfo{author}{\bibfnamefont{B.}~\bibnamefont{{Follin}}},
  \bibinfo{author}{\bibfnamefont{L.}~\bibnamefont{{Knox}}},
  \bibinfo{author}{\bibfnamefont{M.}~\bibnamefont{{Millea}}}, \bibnamefont{and}
  \bibinfo{author}{\bibfnamefont{Z.}~\bibnamefont{{Pan}}},
  \bibinfo{journal}{\prl} \textbf{\bibinfo{volume}{115}}, \bibinfo{eid}{091301}
  (\bibinfo{year}{2015}), \eprint{1503.07863}.

\bibitem[{\citenamefont{{Hu} et~al.}(2001)\citenamefont{{Hu}, {Fukugita},
  {Zaldarriaga}, and {Tegmark}}}]{hu2001}
\bibinfo{author}{\bibfnamefont{W.}~\bibnamefont{{Hu}}},
  \bibinfo{author}{\bibfnamefont{M.}~\bibnamefont{{Fukugita}}},
  \bibinfo{author}{\bibfnamefont{M.}~\bibnamefont{{Zaldarriaga}}},
  \bibnamefont{and}
  \bibinfo{author}{\bibfnamefont{M.}~\bibnamefont{{Tegmark}}},
  \bibinfo{journal}{\apj} \textbf{\bibinfo{volume}{549}}, \bibinfo{pages}{669}
  (\bibinfo{year}{2001}), \eprint{astro-ph/0006436}.

\bibitem[{\citenamefont{Knox and Millea}(2020)}]{hubblehunters}
\bibinfo{author}{\bibfnamefont{L.}~\bibnamefont{Knox}} \bibnamefont{and}
  \bibinfo{author}{\bibfnamefont{M.}~\bibnamefont{Millea}},
  \bibinfo{journal}{Physical Review D} \textbf{\bibinfo{volume}{101}}
  (\bibinfo{year}{2020}), ISSN \bibinfo{issn}{2470-0029},
  \urlprefix\url{http://dx.doi.org/10.1103/PhysRevD.101.043533}.

\bibitem[{\citenamefont{{Poulin} et~al.}(2019)\citenamefont{{Poulin}, {Smith},
  {Karwal}, and {Kamionkowski}}}]{poulin2019}
\bibinfo{author}{\bibfnamefont{V.}~\bibnamefont{{Poulin}}},
  \bibinfo{author}{\bibfnamefont{T.~L.} \bibnamefont{{Smith}}},
  \bibinfo{author}{\bibfnamefont{T.}~\bibnamefont{{Karwal}}}, \bibnamefont{and}
  \bibinfo{author}{\bibfnamefont{M.}~\bibnamefont{{Kamionkowski}}},
  \bibinfo{journal}{\prl} \textbf{\bibinfo{volume}{122}}, \bibinfo{eid}{221301}
  (\bibinfo{year}{2019}), \eprint{1811.04083}.

\bibitem[{\citenamefont{Riess et~al.}(2021)}]{Riess:2021jrx}
\bibinfo{author}{\bibfnamefont{A.~G.} \bibnamefont{Riess}} \bibnamefont{et~al.}
  (\bibinfo{year}{2021}), \eprint{2112.04510}.

\bibitem[{\citenamefont{Aghanim et~al.}(2020{\natexlab{a}})}]{PCP18}
\bibinfo{author}{\bibfnamefont{N.}~\bibnamefont{Aghanim}} \bibnamefont{et~al.}
  (\bibinfo{collaboration}{Planck}), \bibinfo{journal}{Astron. Astrophys.}
  \textbf{\bibinfo{volume}{641}}, \bibinfo{pages}{A6}
  (\bibinfo{year}{2020}{\natexlab{a}}), \bibinfo{note}{[Erratum:
  Astron.Astrophys. 652, C4 (2021)]}, \eprint{1807.06209}.

\bibitem[{\citenamefont{Macaulay et~al.}(2019)\citenamefont{Macaulay, Nichol,
  Bacon, Brout, Davis, Zhang, Bassett, Scolnic, Möller, D’Andrea
  et~al.}}]{inv-dist-ladder}
\bibinfo{author}{\bibfnamefont{E.}~\bibnamefont{Macaulay}},
  \bibinfo{author}{\bibfnamefont{R.~C.} \bibnamefont{Nichol}},
  \bibinfo{author}{\bibfnamefont{D.}~\bibnamefont{Bacon}},
  \bibinfo{author}{\bibfnamefont{D.}~\bibnamefont{Brout}},
  \bibinfo{author}{\bibfnamefont{T.~M.} \bibnamefont{Davis}},
  \bibinfo{author}{\bibfnamefont{B.}~\bibnamefont{Zhang}},
  \bibinfo{author}{\bibfnamefont{B.}~\bibnamefont{Bassett}},
  \bibinfo{author}{\bibfnamefont{D.}~\bibnamefont{Scolnic}},
  \bibinfo{author}{\bibfnamefont{A.}~\bibnamefont{Möller}},
  \bibinfo{author}{\bibfnamefont{C.~B.} \bibnamefont{D’Andrea}},
  \bibnamefont{et~al.}, \bibinfo{journal}{Monthly Notices of the Royal
  Astronomical Society} \textbf{\bibinfo{volume}{486}}, \bibinfo{pages}{2184}
  (\bibinfo{year}{2019}), ISSN \bibinfo{issn}{0035-8711},
  \eprint{https://academic.oup.com/mnras/article-pdf/486/2/2184/28488227/stz978.pdf},
  \urlprefix\url{https://doi.org/10.1093/mnras/stz978}.

\bibitem[{\citenamefont{Alam et~al.}(2021)\citenamefont{Alam, Aubert, Avila,
  Balland, Bautista, Bershady, Bizyaev, Blanton, Bolton, Bovy
  et~al.}}]{BOSS-EBOSS-inv-dist-ladder}
\bibinfo{author}{\bibfnamefont{S.}~\bibnamefont{Alam}},
  \bibinfo{author}{\bibfnamefont{M.}~\bibnamefont{Aubert}},
  \bibinfo{author}{\bibfnamefont{S.}~\bibnamefont{Avila}},
  \bibinfo{author}{\bibfnamefont{C.}~\bibnamefont{Balland}},
  \bibinfo{author}{\bibfnamefont{J.~E.} \bibnamefont{Bautista}},
  \bibinfo{author}{\bibfnamefont{M.~A.} \bibnamefont{Bershady}},
  \bibinfo{author}{\bibfnamefont{D.}~\bibnamefont{Bizyaev}},
  \bibinfo{author}{\bibfnamefont{M.~R.} \bibnamefont{Blanton}},
  \bibinfo{author}{\bibfnamefont{A.~S.} \bibnamefont{Bolton}},
  \bibinfo{author}{\bibfnamefont{J.}~\bibnamefont{Bovy}}, \bibnamefont{et~al.},
  \bibinfo{journal}{Phys. Rev. D} \textbf{\bibinfo{volume}{103}},
  \bibinfo{pages}{083533} (\bibinfo{year}{2021}),
  \urlprefix\url{https://link.aps.org/doi/10.1103/PhysRevD.103.083533}.

\bibitem[{\citenamefont{Addison et~al.}(2018)\citenamefont{Addison, Watts,
  Bennett, Halpern, Hinshaw, and Weiland}}]{Addison_2018-inv-dist-ladder}
\bibinfo{author}{\bibfnamefont{G.~E.} \bibnamefont{Addison}},
  \bibinfo{author}{\bibfnamefont{D.~J.} \bibnamefont{Watts}},
  \bibinfo{author}{\bibfnamefont{C.~L.} \bibnamefont{Bennett}},
  \bibinfo{author}{\bibfnamefont{M.}~\bibnamefont{Halpern}},
  \bibinfo{author}{\bibfnamefont{G.}~\bibnamefont{Hinshaw}}, \bibnamefont{and}
  \bibinfo{author}{\bibfnamefont{J.~L.} \bibnamefont{Weiland}},
  \bibinfo{journal}{The Astrophysical Journal} \textbf{\bibinfo{volume}{853}},
  \bibinfo{pages}{119} (\bibinfo{year}{2018}),
  \urlprefix\url{https://doi.org/10.3847%2F1538-4357%2Faaa1ed}.

\bibitem[{\citenamefont{Abbott et~al.}(2018)\citenamefont{Abbott, Abdalla,
  Annis, Bechtol, Blazek, Benson, Bernstein, Bernstein, Bertin, Brooks
  et~al.}}]{BAO-BBN-LSS}
\bibinfo{author}{\bibfnamefont{T.~M.~C.} \bibnamefont{Abbott}},
  \bibinfo{author}{\bibfnamefont{F.~B.} \bibnamefont{Abdalla}},
  \bibinfo{author}{\bibfnamefont{J.}~\bibnamefont{Annis}},
  \bibinfo{author}{\bibfnamefont{K.}~\bibnamefont{Bechtol}},
  \bibinfo{author}{\bibfnamefont{J.}~\bibnamefont{Blazek}},
  \bibinfo{author}{\bibfnamefont{B.~A.} \bibnamefont{Benson}},
  \bibinfo{author}{\bibfnamefont{R.~A.} \bibnamefont{Bernstein}},
  \bibinfo{author}{\bibfnamefont{G.~M.} \bibnamefont{Bernstein}},
  \bibinfo{author}{\bibfnamefont{E.}~\bibnamefont{Bertin}},
  \bibinfo{author}{\bibfnamefont{D.}~\bibnamefont{Brooks}},
  \bibnamefont{et~al.}, \bibinfo{journal}{Monthly Notices of the Royal
  Astronomical Society} \textbf{\bibinfo{volume}{480}}, \bibinfo{pages}{3879}
  (\bibinfo{year}{2018}),
  \urlprefix\url{https://doi.org/10.1093%2Fmnras%2Fsty1939}.

\bibitem[{\citenamefont{Cuceu et~al.}(2019)\citenamefont{Cuceu, Farr, Lemos,
  and Font-Ribera}}]{BAO-BBN-2}
\bibinfo{author}{\bibfnamefont{A.}~\bibnamefont{Cuceu}},
  \bibinfo{author}{\bibfnamefont{J.}~\bibnamefont{Farr}},
  \bibinfo{author}{\bibfnamefont{P.}~\bibnamefont{Lemos}}, \bibnamefont{and}
  \bibinfo{author}{\bibfnamefont{A.}~\bibnamefont{Font-Ribera}},
  \bibinfo{journal}{Journal of Cosmology and Astroparticle Physics}
  \textbf{\bibinfo{volume}{2019}}, \bibinfo{pages}{044} (\bibinfo{year}{2019}),
  \urlprefix\url{https://doi.org/10.1088/1475-7516/2019/10/044}.

\bibitem[{\citenamefont{Schöneberg et~al.}(2019)\citenamefont{Schöneberg,
  Lesgourgues, and Hooper}}]{BAO-BBN}
\bibinfo{author}{\bibfnamefont{N.}~\bibnamefont{Schöneberg}},
  \bibinfo{author}{\bibfnamefont{J.}~\bibnamefont{Lesgourgues}},
  \bibnamefont{and} \bibinfo{author}{\bibfnamefont{D.~C.}
  \bibnamefont{Hooper}}, \bibinfo{journal}{Journal of Cosmology and
  Astroparticle Physics} \textbf{\bibinfo{volume}{2019}}, \bibinfo{pages}{029}
  (\bibinfo{year}{2019}),
  \urlprefix\url{https://doi.org/10.1088%2F1475-7516%2F2019%2F10%2F029}.

\bibitem[{\citenamefont{{Freedman} et~al.}(2019)\citenamefont{{Freedman},
  {Madore}, {Hatt}, {Hoyt}, {Jang}, {Beaton}, {Burns}, {Lee}, {Monson},
  {Neeley} et~al.}}]{freedman:2019}
\bibinfo{author}{\bibfnamefont{W.~L.} \bibnamefont{{Freedman}}},
  \bibinfo{author}{\bibfnamefont{B.~F.} \bibnamefont{{Madore}}},
  \bibinfo{author}{\bibfnamefont{D.}~\bibnamefont{{Hatt}}},
  \bibinfo{author}{\bibfnamefont{T.~J.} \bibnamefont{{Hoyt}}},
  \bibinfo{author}{\bibfnamefont{I.~S.} \bibnamefont{{Jang}}},
  \bibinfo{author}{\bibfnamefont{R.~L.} \bibnamefont{{Beaton}}},
  \bibinfo{author}{\bibfnamefont{C.~R.} \bibnamefont{{Burns}}},
  \bibinfo{author}{\bibfnamefont{M.~G.} \bibnamefont{{Lee}}},
  \bibinfo{author}{\bibfnamefont{A.~J.} \bibnamefont{{Monson}}},
  \bibinfo{author}{\bibfnamefont{J.~R.} \bibnamefont{{Neeley}}},
  \bibnamefont{et~al.}, \bibinfo{journal}{\apj} \textbf{\bibinfo{volume}{882}},
  \bibinfo{eid}{34} (\bibinfo{year}{2019}), \eprint{1907.05922}.

\bibitem[{\citenamefont{Shah et~al.}(2021)\citenamefont{Shah, Lemos, and
  Lahav}}]{Shah:2021}
\bibinfo{author}{\bibfnamefont{P.}~\bibnamefont{Shah}},
  \bibinfo{author}{\bibfnamefont{P.}~\bibnamefont{Lemos}}, \bibnamefont{and}
  \bibinfo{author}{\bibfnamefont{O.}~\bibnamefont{Lahav}},
  \bibinfo{journal}{Astron. Astrophys. Rev.} \textbf{\bibinfo{volume}{29}},
  \bibinfo{pages}{9} (\bibinfo{year}{2021}), \eprint{2109.01161}.

\bibitem[{\citenamefont{Verde et~al.}(2019)\citenamefont{Verde, Treu, and
  Riess}}]{Verde19}
\bibinfo{author}{\bibfnamefont{L.}~\bibnamefont{Verde}},
  \bibinfo{author}{\bibfnamefont{T.}~\bibnamefont{Treu}}, \bibnamefont{and}
  \bibinfo{author}{\bibfnamefont{A.~G.} \bibnamefont{Riess}},
  \bibinfo{journal}{Nature Astronomy} \textbf{\bibinfo{volume}{3}},
  \bibinfo{pages}{891–895} (\bibinfo{year}{2019}), ISSN
  \bibinfo{issn}{2397-3366},
  \urlprefix\url{http://dx.doi.org/10.1038/s41550-019-0902-0}.

\bibitem[{\citenamefont{Riess}(2019)}]{Riess19-2}
\bibinfo{author}{\bibfnamefont{A.~G.} \bibnamefont{Riess}},
  \bibinfo{journal}{Nature Reviews Physics} \textbf{\bibinfo{volume}{2}},
  \bibinfo{pages}{10–12} (\bibinfo{year}{2019}), ISSN
  \bibinfo{issn}{2522-5820},
  \urlprefix\url{http://dx.doi.org/10.1038/s42254-019-0137-0}.

\bibitem[{\citenamefont{Di~Valentino
  et~al.}(2021{\natexlab{a}})\citenamefont{Di~Valentino, Mena, Pan, Visinelli,
  Yang, Melchiorri, Mota, Riess, and Silk}}]{Valentino-review-solns}
\bibinfo{author}{\bibfnamefont{E.}~\bibnamefont{Di~Valentino}},
  \bibinfo{author}{\bibfnamefont{O.}~\bibnamefont{Mena}},
  \bibinfo{author}{\bibfnamefont{S.}~\bibnamefont{Pan}},
  \bibinfo{author}{\bibfnamefont{L.}~\bibnamefont{Visinelli}},
  \bibinfo{author}{\bibfnamefont{W.}~\bibnamefont{Yang}},
  \bibinfo{author}{\bibfnamefont{A.}~\bibnamefont{Melchiorri}},
  \bibinfo{author}{\bibfnamefont{D.~F.} \bibnamefont{Mota}},
  \bibinfo{author}{\bibfnamefont{A.~G.} \bibnamefont{Riess}}, \bibnamefont{and}
  \bibinfo{author}{\bibfnamefont{J.}~\bibnamefont{Silk}},
  \bibinfo{journal}{Classical and Quantum Gravity}
  \textbf{\bibinfo{volume}{38}}, \bibinfo{pages}{153001}
  (\bibinfo{year}{2021}{\natexlab{a}}), ISSN \bibinfo{issn}{1361-6382},
  \urlprefix\url{http://dx.doi.org/10.1088/1361-6382/ac086d}.

\bibitem[{\citenamefont{Sch\"oneberg et~al.}(2021)\citenamefont{Sch\"oneberg,
  Franco~Abell\'an, P\'erez~S\'anchez, Witte, Poulin, and
  Lesgourgues}}]{h0olympics}
\bibinfo{author}{\bibfnamefont{N.}~\bibnamefont{Sch\"oneberg}},
  \bibinfo{author}{\bibfnamefont{G.}~\bibnamefont{Franco~Abell\'an}},
  \bibinfo{author}{\bibfnamefont{A.}~\bibnamefont{P\'erez~S\'anchez}},
  \bibinfo{author}{\bibfnamefont{S.~J.} \bibnamefont{Witte}},
  \bibinfo{author}{\bibfnamefont{V.}~\bibnamefont{Poulin}}, \bibnamefont{and}
  \bibinfo{author}{\bibfnamefont{J.}~\bibnamefont{Lesgourgues}}
  (\bibinfo{year}{2021}), \eprint{2107.10291}.

\bibitem[{\citenamefont{Asgari et~al.}(2021)\citenamefont{Asgari, Lin,
  Joachimi, Giblin, Heymans, Hildebrandt, Kannawadi, Stölzner, Tröster,
  van~den Busch et~al.}}]{Asgari_2021}
\bibinfo{author}{\bibfnamefont{M.}~\bibnamefont{Asgari}},
  \bibinfo{author}{\bibfnamefont{C.-A.} \bibnamefont{Lin}},
  \bibinfo{author}{\bibfnamefont{B.}~\bibnamefont{Joachimi}},
  \bibinfo{author}{\bibfnamefont{B.}~\bibnamefont{Giblin}},
  \bibinfo{author}{\bibfnamefont{C.}~\bibnamefont{Heymans}},
  \bibinfo{author}{\bibfnamefont{H.}~\bibnamefont{Hildebrandt}},
  \bibinfo{author}{\bibfnamefont{A.}~\bibnamefont{Kannawadi}},
  \bibinfo{author}{\bibfnamefont{B.}~\bibnamefont{Stölzner}},
  \bibinfo{author}{\bibfnamefont{T.}~\bibnamefont{Tröster}},
  \bibinfo{author}{\bibfnamefont{J.~L.} \bibnamefont{van~den Busch}},
  \bibnamefont{et~al.}, \bibinfo{journal}{Astronomy \& Astrophysics}
  \textbf{\bibinfo{volume}{645}}, \bibinfo{pages}{A104} (\bibinfo{year}{2021}),
  ISSN \bibinfo{issn}{1432-0746},
  \urlprefix\url{http://dx.doi.org/10.1051/0004-6361/202039070}.

\bibitem[{\citenamefont{Heymans et~al.}(2021)\citenamefont{Heymans, Tröster,
  Asgari, Blake, Hildebrandt, Joachimi, Kuijken, Lin, Sánchez, van~den Busch
  et~al.}}]{Heymans_2021}
\bibinfo{author}{\bibfnamefont{C.}~\bibnamefont{Heymans}},
  \bibinfo{author}{\bibfnamefont{T.}~\bibnamefont{Tröster}},
  \bibinfo{author}{\bibfnamefont{M.}~\bibnamefont{Asgari}},
  \bibinfo{author}{\bibfnamefont{C.}~\bibnamefont{Blake}},
  \bibinfo{author}{\bibfnamefont{H.}~\bibnamefont{Hildebrandt}},
  \bibinfo{author}{\bibfnamefont{B.}~\bibnamefont{Joachimi}},
  \bibinfo{author}{\bibfnamefont{K.}~\bibnamefont{Kuijken}},
  \bibinfo{author}{\bibfnamefont{C.-A.} \bibnamefont{Lin}},
  \bibinfo{author}{\bibfnamefont{A.~G.} \bibnamefont{Sánchez}},
  \bibinfo{author}{\bibfnamefont{J.~L.} \bibnamefont{van~den Busch}},
  \bibnamefont{et~al.}, \bibinfo{journal}{Astronomy \& Astrophysics}
  \textbf{\bibinfo{volume}{646}}, \bibinfo{pages}{A140} (\bibinfo{year}{2021}),
  ISSN \bibinfo{issn}{1432-0746},
  \urlprefix\url{http://dx.doi.org/10.1051/0004-6361/202039063}.

\bibitem[{\citenamefont{Di~Valentino
  et~al.}(2021{\natexlab{b}})}]{cosmology-intertwined-III}
\bibinfo{author}{\bibfnamefont{E.}~\bibnamefont{Di~Valentino}}
  \bibnamefont{et~al.}, \bibinfo{journal}{Astropart. Phys.}
  \textbf{\bibinfo{volume}{131}}, \bibinfo{pages}{102604}
  (\bibinfo{year}{2021}{\natexlab{b}}), \eprint{2008.11285}.

\bibitem[{\citenamefont{Abdalla et~al.}(2022)}]{Abdalla:2022yfr}
\bibinfo{author}{\bibfnamefont{E.}~\bibnamefont{Abdalla}} \bibnamefont{et~al.},
  \bibinfo{journal}{JHEAp} \textbf{\bibinfo{volume}{34}}, \bibinfo{pages}{49}
  (\bibinfo{year}{2022}), \eprint{2203.06142}.

\bibitem[{\citenamefont{Hill et~al.}(2020)\citenamefont{Hill, McDonough,
  Toomey, and Alexander}}]{EDE-Hill-LSS}
\bibinfo{author}{\bibfnamefont{J.~C.} \bibnamefont{Hill}},
  \bibinfo{author}{\bibfnamefont{E.}~\bibnamefont{McDonough}},
  \bibinfo{author}{\bibfnamefont{M.~W.} \bibnamefont{Toomey}},
  \bibnamefont{and}
  \bibinfo{author}{\bibfnamefont{S.}~\bibnamefont{Alexander}},
  \bibinfo{journal}{Physical Review D} \textbf{\bibinfo{volume}{102}}
  (\bibinfo{year}{2020}), ISSN \bibinfo{issn}{2470-0029},
  \urlprefix\url{http://dx.doi.org/10.1103/PhysRevD.102.043507}.

\bibitem[{\citenamefont{Ivanov et~al.}(2020)\citenamefont{Ivanov, McDonough,
  Hill, Simonović, Toomey, Alexander, and Zaldarriaga}}]{EDE-Ivanov-LSS}
\bibinfo{author}{\bibfnamefont{M.~M.} \bibnamefont{Ivanov}},
  \bibinfo{author}{\bibfnamefont{E.}~\bibnamefont{McDonough}},
  \bibinfo{author}{\bibfnamefont{J.~C.} \bibnamefont{Hill}},
  \bibinfo{author}{\bibfnamefont{M.}~\bibnamefont{Simonović}},
  \bibinfo{author}{\bibfnamefont{M.~W.} \bibnamefont{Toomey}},
  \bibinfo{author}{\bibfnamefont{S.}~\bibnamefont{Alexander}},
  \bibnamefont{and}
  \bibinfo{author}{\bibfnamefont{M.}~\bibnamefont{Zaldarriaga}},
  \bibinfo{journal}{Physical Review D} \textbf{\bibinfo{volume}{102}}
  (\bibinfo{year}{2020}), ISSN \bibinfo{issn}{2470-0029},
  \urlprefix\url{http://dx.doi.org/10.1103/PhysRevD.102.103502}.

\bibitem[{\citenamefont{Tröster et~al.}(2021)\citenamefont{Tröster, Asgari,
  Blake, Cataneo, Heymans, Hildebrandt, Joachimi, Lin, Sánchez, Wright
  et~al.}}]{Troster-S8-beyond-LCDM}
\bibinfo{author}{\bibfnamefont{T.}~\bibnamefont{Tröster}},
  \bibinfo{author}{\bibfnamefont{M.}~\bibnamefont{Asgari}},
  \bibinfo{author}{\bibfnamefont{C.}~\bibnamefont{Blake}},
  \bibinfo{author}{\bibfnamefont{M.}~\bibnamefont{Cataneo}},
  \bibinfo{author}{\bibfnamefont{C.}~\bibnamefont{Heymans}},
  \bibinfo{author}{\bibfnamefont{H.}~\bibnamefont{Hildebrandt}},
  \bibinfo{author}{\bibfnamefont{B.}~\bibnamefont{Joachimi}},
  \bibinfo{author}{\bibfnamefont{C.-A.} \bibnamefont{Lin}},
  \bibinfo{author}{\bibfnamefont{A.~G.} \bibnamefont{Sánchez}},
  \bibinfo{author}{\bibfnamefont{A.~H.} \bibnamefont{Wright}},
  \bibnamefont{et~al.}, \bibinfo{journal}{Astronomy \& Astrophysics}
  \textbf{\bibinfo{volume}{649}}, \bibinfo{pages}{A88} (\bibinfo{year}{2021}),
  ISSN \bibinfo{issn}{1432-0746},
  \urlprefix\url{http://dx.doi.org/10.1051/0004-6361/202039805}.

\bibitem[{\citenamefont{Roy et~al.}(2018)\citenamefont{Roy, Gonzalez-Morales,
  and Urena-Lopez}}]{Roy-no-perturbations}
\bibinfo{author}{\bibfnamefont{N.}~\bibnamefont{Roy}},
  \bibinfo{author}{\bibfnamefont{A.~X.} \bibnamefont{Gonzalez-Morales}},
  \bibnamefont{and} \bibinfo{author}{\bibfnamefont{L.~A.}
  \bibnamefont{Urena-Lopez}}, \bibinfo{journal}{Phys. Rev. D}
  \textbf{\bibinfo{volume}{98}}, \bibinfo{pages}{063530}
  (\bibinfo{year}{2018}), \eprint{1803.09204}.

\bibitem[{\citenamefont{Ure\~na L\'opez and
  Roy}(2020)}]{Roy-with-perturbations}
\bibinfo{author}{\bibfnamefont{L.~A.} \bibnamefont{Ure\~na L\'opez}}
  \bibnamefont{and} \bibinfo{author}{\bibfnamefont{N.}~\bibnamefont{Roy}},
  \bibinfo{journal}{Phys. Rev. D} \textbf{\bibinfo{volume}{102}},
  \bibinfo{pages}{063510} (\bibinfo{year}{2020}), \eprint{2007.08873}.

\bibitem[{\citenamefont{Caldwell et~al.}(1998)\citenamefont{Caldwell, Dave, and
  Steinhardt}}]{Caldwell-quintessence-perturbations-imprint}
\bibinfo{author}{\bibfnamefont{R.~R.} \bibnamefont{Caldwell}},
  \bibinfo{author}{\bibfnamefont{R.}~\bibnamefont{Dave}}, \bibnamefont{and}
  \bibinfo{author}{\bibfnamefont{P.~J.} \bibnamefont{Steinhardt}},
  \bibinfo{journal}{Phys. Rev. Lett.} \textbf{\bibinfo{volume}{80}},
  \bibinfo{pages}{1582} (\bibinfo{year}{1998}), \eprint{astro-ph/9708069}.

\bibitem[{\citenamefont{Albrecht
  et~al.}(2002{\natexlab{a}})\citenamefont{Albrecht, Burgess, Ravndal, and
  Skordis}}]{Albrecht:2001cp}
\bibinfo{author}{\bibfnamefont{A.}~\bibnamefont{Albrecht}},
  \bibinfo{author}{\bibfnamefont{C.~P.} \bibnamefont{Burgess}},
  \bibinfo{author}{\bibfnamefont{F.}~\bibnamefont{Ravndal}}, \bibnamefont{and}
  \bibinfo{author}{\bibfnamefont{C.}~\bibnamefont{Skordis}},
  \bibinfo{journal}{Phys. Rev. D} \textbf{\bibinfo{volume}{65}},
  \bibinfo{pages}{123506} (\bibinfo{year}{2002}{\natexlab{a}}),
  \eprint{hep-th/0105261}.

\bibitem[{\citenamefont{Albrecht
  et~al.}(2002{\natexlab{b}})\citenamefont{Albrecht, Burgess, Ravndal, and
  Skordis}}]{Albrecht:2001xt}
\bibinfo{author}{\bibfnamefont{A.}~\bibnamefont{Albrecht}},
  \bibinfo{author}{\bibfnamefont{C.~P.} \bibnamefont{Burgess}},
  \bibinfo{author}{\bibfnamefont{F.}~\bibnamefont{Ravndal}}, \bibnamefont{and}
  \bibinfo{author}{\bibfnamefont{C.}~\bibnamefont{Skordis}},
  \bibinfo{journal}{Phys. Rev. D} \textbf{\bibinfo{volume}{65}},
  \bibinfo{pages}{123507} (\bibinfo{year}{2002}{\natexlab{b}}),
  \eprint{astro-ph/0107573}.

\bibitem[{\citenamefont{Ferreira and Joyce}(1998)}]{FerJoyce}
\bibinfo{author}{\bibfnamefont{P.~G.} \bibnamefont{Ferreira}} \bibnamefont{and}
  \bibinfo{author}{\bibfnamefont{M.}~\bibnamefont{Joyce}},
  \bibinfo{journal}{Physical Review D} \textbf{\bibinfo{volume}{58}}
  (\bibinfo{year}{1998}), ISSN \bibinfo{issn}{1089-4918},
  \urlprefix\url{http://dx.doi.org/10.1103/PhysRevD.58.023503}.

\bibitem[{\citenamefont{{Bashinsky} and {Seljak}}(2004)}]{bashinsky04}
\bibinfo{author}{\bibfnamefont{S.}~\bibnamefont{{Bashinsky}}} \bibnamefont{and}
  \bibinfo{author}{\bibfnamefont{U.}~\bibnamefont{{Seljak}}},
  \bibinfo{journal}{\prd} \textbf{\bibinfo{volume}{69}}, \bibinfo{eid}{083002}
  (\bibinfo{year}{2004}), \eprint{astro-ph/0310198}.

\bibitem[{\citenamefont{Pan et~al.}(2016)\citenamefont{Pan, Knox, Mulroe, and
  Narimani}}]{pan16}
\bibinfo{author}{\bibfnamefont{Z.}~\bibnamefont{Pan}},
  \bibinfo{author}{\bibfnamefont{L.}~\bibnamefont{Knox}},
  \bibinfo{author}{\bibfnamefont{B.}~\bibnamefont{Mulroe}}, \bibnamefont{and}
  \bibinfo{author}{\bibfnamefont{A.}~\bibnamefont{Narimani}},
  \bibinfo{journal}{Monthly Notices of the Royal Astronomical Society}
  \textbf{\bibinfo{volume}{459}}, \bibinfo{pages}{2513} (\bibinfo{year}{2016}).

\bibitem[{\citenamefont{Ballesteros and
  Lesgourgues}(2010)}]{Lesgourgues_and_Ballesteros}
\bibinfo{author}{\bibfnamefont{G.}~\bibnamefont{Ballesteros}} \bibnamefont{and}
  \bibinfo{author}{\bibfnamefont{J.}~\bibnamefont{Lesgourgues}},
  \bibinfo{journal}{Journal of Cosmology and Astroparticle Physics}
  \textbf{\bibinfo{volume}{2010}}, \bibinfo{pages}{014–014}
  (\bibinfo{year}{2010}), ISSN \bibinfo{issn}{1475-7516},
  \urlprefix\url{http://dx.doi.org/10.1088/1475-7516/2010/10/014}.

\bibitem[{\citenamefont{Jennings et~al.}(2010)\citenamefont{Jennings, Baugh,
  Angulo, and Pascoli}}]{Jennings_2010}
\bibinfo{author}{\bibfnamefont{E.}~\bibnamefont{Jennings}},
  \bibinfo{author}{\bibfnamefont{C.~M.} \bibnamefont{Baugh}},
  \bibinfo{author}{\bibfnamefont{R.~E.} \bibnamefont{Angulo}},
  \bibnamefont{and} \bibinfo{author}{\bibfnamefont{S.}~\bibnamefont{Pascoli}},
  \bibinfo{journal}{Monthly Notices of the Royal Astronomical Society}
  \textbf{\bibinfo{volume}{401}}, \bibinfo{pages}{2181–2201}
  (\bibinfo{year}{2010}), ISSN \bibinfo{issn}{1365-2966},
  \urlprefix\url{http://dx.doi.org/10.1111/j.1365-2966.2009.15819.x}.

\bibitem[{\citenamefont{Brax et~al.}(2000)\citenamefont{Brax, Martin, and
  Riazuelo}}]{Jerome-Martin-P-Brax}
\bibinfo{author}{\bibfnamefont{P.}~\bibnamefont{Brax}},
  \bibinfo{author}{\bibfnamefont{J.}~\bibnamefont{Martin}}, \bibnamefont{and}
  \bibinfo{author}{\bibfnamefont{A.}~\bibnamefont{Riazuelo}},
  \bibinfo{journal}{Phys. Rev. D} \textbf{\bibinfo{volume}{62}},
  \bibinfo{pages}{103505} (\bibinfo{year}{2000}), \eprint{astro-ph/0005428}.

\bibitem[{\citenamefont{Hu and Sugiyama}(1995)}]{Hu-Sugiyama-Rad-Driving}
\bibinfo{author}{\bibfnamefont{W.}~\bibnamefont{Hu}} \bibnamefont{and}
  \bibinfo{author}{\bibfnamefont{N.}~\bibnamefont{Sugiyama}},
  \bibinfo{journal}{The Astrophysical Journal} \textbf{\bibinfo{volume}{471}}
  (\bibinfo{year}{1995}).

\bibitem[{\citenamefont{Alam et~al.}(2017)}]{Alam-BOSS-BAO-DR12}
\bibinfo{author}{\bibfnamefont{S.}~\bibnamefont{Alam}} \bibnamefont{et~al.}
  (\bibinfo{collaboration}{BOSS}), \bibinfo{journal}{Mon. Not. Roy. Astron.
  Soc.} \textbf{\bibinfo{volume}{470}}, \bibinfo{pages}{2617}
  (\bibinfo{year}{2017}), \eprint{1607.03155}.

\bibitem[{\citenamefont{Sherwin and
  White}(2019)}]{White-Sherwin-BAO-Reconstruction-Errors}
\bibinfo{author}{\bibfnamefont{B.~D.} \bibnamefont{Sherwin}} \bibnamefont{and}
  \bibinfo{author}{\bibfnamefont{M.}~\bibnamefont{White}},
  \bibinfo{journal}{JCAP} \textbf{\bibinfo{volume}{02}}, \bibinfo{pages}{027}
  (\bibinfo{year}{2019}), \eprint{1808.04384}.

\bibitem[{\citenamefont{Aghanim
  et~al.}(2020{\natexlab{b}})}]{Planck-Collab-likelihoods}
\bibinfo{author}{\bibfnamefont{N.}~\bibnamefont{Aghanim}} \bibnamefont{et~al.}
  (\bibinfo{collaboration}{Planck}), \bibinfo{journal}{Astron. Astrophys.}
  \textbf{\bibinfo{volume}{641}}, \bibinfo{pages}{A5}
  (\bibinfo{year}{2020}{\natexlab{b}}), \eprint{1907.12875}.

\bibitem[{\citenamefont{Aghanim
  et~al.}(2020{\natexlab{c}})}]{Planck-Collab-lensing}
\bibinfo{author}{\bibfnamefont{N.}~\bibnamefont{Aghanim}} \bibnamefont{et~al.}
  (\bibinfo{collaboration}{Planck}), \bibinfo{journal}{Astron. Astrophys.}
  \textbf{\bibinfo{volume}{641}}, \bibinfo{pages}{A8}
  (\bibinfo{year}{2020}{\natexlab{c}}), \eprint{1807.06210}.

\bibitem[{\citenamefont{Scolnic et~al.}(2018)}]{Pantheon}
\bibinfo{author}{\bibfnamefont{D.~M.} \bibnamefont{Scolnic}}
  \bibnamefont{et~al.} (\bibinfo{collaboration}{Pan-STARRS1}),
  \bibinfo{journal}{Astrophys. J.} \textbf{\bibinfo{volume}{859}},
  \bibinfo{pages}{101} (\bibinfo{year}{2018}), \eprint{1710.00845}.

\bibitem[{\citenamefont{Riess et~al.}(2019)\citenamefont{Riess, Casertano,
  Yuan, Macri, and Scolnic}}]{Riess19}
\bibinfo{author}{\bibfnamefont{A.~G.} \bibnamefont{Riess}},
  \bibinfo{author}{\bibfnamefont{S.}~\bibnamefont{Casertano}},
  \bibinfo{author}{\bibfnamefont{W.}~\bibnamefont{Yuan}},
  \bibinfo{author}{\bibfnamefont{L.~M.} \bibnamefont{Macri}}, \bibnamefont{and}
  \bibinfo{author}{\bibfnamefont{D.}~\bibnamefont{Scolnic}},
  \bibinfo{journal}{The Astrophysical Journal} \textbf{\bibinfo{volume}{876}},
  \bibinfo{pages}{85} (\bibinfo{year}{2019}), ISSN \bibinfo{issn}{1538-4357},
  \urlprefix\url{http://dx.doi.org/10.3847/1538-4357/ab1422}.

\bibitem[{\citenamefont{Ade et~al.}(2014)}]{Planck-SZ}
\bibinfo{author}{\bibfnamefont{P.~A.~R.} \bibnamefont{Ade}}
  \bibnamefont{et~al.} (\bibinfo{collaboration}{Planck}),
  \bibinfo{journal}{Astron. Astrophys.} \textbf{\bibinfo{volume}{571}},
  \bibinfo{pages}{A20} (\bibinfo{year}{2014}), \eprint{1303.5080}.

\bibitem[{\citenamefont{Joudaki et~al.}(2020)}]{Joudaki-S8-prior}
\bibinfo{author}{\bibfnamefont{S.}~\bibnamefont{Joudaki}} \bibnamefont{et~al.},
  \bibinfo{journal}{Astron. Astrophys.} \textbf{\bibinfo{volume}{638}},
  \bibinfo{pages}{L1} (\bibinfo{year}{2020}), \eprint{1906.09262}.

\bibitem[{\citenamefont{"Troxel and others"}(2018)}]{Troxel-18-S8-prior}
\bibinfo{author}{\bibfnamefont{M.~A.} \bibnamefont{"Troxel}} \bibnamefont{and}
  \bibinfo{author}{\bibnamefont{others"}}, \bibinfo{journal}{Phys. Rev. D}
  \textbf{\bibinfo{volume}{98}}, \bibinfo{pages}{043528}
  (\bibinfo{year}{2018}),
  \urlprefix\url{https://link.aps.org/doi/10.1103/PhysRevD.98.043528}.

\bibitem[{\citenamefont{Coleman and De~Luccia}(1980)}]{Coleman-DeLuccia}
\bibinfo{author}{\bibfnamefont{S.}~\bibnamefont{Coleman}} \bibnamefont{and}
  \bibinfo{author}{\bibfnamefont{F.}~\bibnamefont{De~Luccia}},
  \bibinfo{journal}{Phys. Rev. D} \textbf{\bibinfo{volume}{21}},
  \bibinfo{pages}{3305} (\bibinfo{year}{1980}),
  \urlprefix\url{https://link.aps.org/doi/10.1103/PhysRevD.21.3305}.

\bibitem[{\citenamefont{WELLER}(2000)}]{Bubbles-Jochen-Weller}
\bibinfo{author}{\bibfnamefont{J.}~\bibnamefont{WELLER}},
  \bibinfo{journal}{Particles, Strings and Cosmology (PASCOS 99)}
  (\bibinfo{year}{2000}),
  \urlprefix\url{http://dx.doi.org/10.1142/9789812792433_0051}.

\bibitem[{\citenamefont{Barrow et~al.}(2000)\citenamefont{Barrow, Bean, and
  Magueijo}}]{10.1046/j.1365-8711.2000.03778.x}
\bibinfo{author}{\bibfnamefont{J.~D.} \bibnamefont{Barrow}},
  \bibinfo{author}{\bibfnamefont{R.}~\bibnamefont{Bean}}, \bibnamefont{and}
  \bibinfo{author}{\bibfnamefont{J.}~\bibnamefont{Magueijo}},
  \bibinfo{journal}{Monthly Notices of the Royal Astronomical Society}
  \textbf{\bibinfo{volume}{316}}, \bibinfo{pages}{L41} (\bibinfo{year}{2000}),
  ISSN \bibinfo{issn}{0035-8711}.

\bibitem[{\citenamefont{Blas et~al.}(2011)\citenamefont{Blas, Lesgourgues, and
  Tram}}]{CLASS-paper}
\bibinfo{author}{\bibfnamefont{D.}~\bibnamefont{Blas}},
  \bibinfo{author}{\bibfnamefont{J.}~\bibnamefont{Lesgourgues}},
  \bibnamefont{and} \bibinfo{author}{\bibfnamefont{T.}~\bibnamefont{Tram}},
  \bibinfo{journal}{Journal of Cosmology and Astroparticle Physics}
  \textbf{\bibinfo{volume}{2011}}, \bibinfo{pages}{034–034}
  (\bibinfo{year}{2011}), ISSN \bibinfo{issn}{1475-7516},
  \urlprefix\url{http://dx.doi.org/10.1088/1475-7516/2011/07/034}.

\bibitem[{\citenamefont{Brinckmann and Lesgourgues}(2018)}]{MontePython1}
\bibinfo{author}{\bibfnamefont{T.}~\bibnamefont{Brinckmann}} \bibnamefont{and}
  \bibinfo{author}{\bibfnamefont{J.}~\bibnamefont{Lesgourgues}}
  (\bibinfo{year}{2018}), \eprint{1804.07261}.

\bibitem[{\citenamefont{Audren et~al.}(2013)\citenamefont{Audren, Lesgourgues,
  Benabed, and Prunet}}]{MontePython2}
\bibinfo{author}{\bibfnamefont{B.}~\bibnamefont{Audren}},
  \bibinfo{author}{\bibfnamefont{J.}~\bibnamefont{Lesgourgues}},
  \bibinfo{author}{\bibfnamefont{K.}~\bibnamefont{Benabed}}, \bibnamefont{and}
  \bibinfo{author}{\bibfnamefont{S.}~\bibnamefont{Prunet}},
  \bibinfo{journal}{JCAP} \textbf{\bibinfo{volume}{1302}}, \bibinfo{pages}{001}
  (\bibinfo{year}{2013}), \eprint{1210.7183}.

\bibitem[{\citenamefont{Feroz and Hobson}(2008)}]{MultiNest1}
\bibinfo{author}{\bibfnamefont{F.}~\bibnamefont{Feroz}} \bibnamefont{and}
  \bibinfo{author}{\bibfnamefont{M.~P.} \bibnamefont{Hobson}},
  \bibinfo{journal}{Monthly Notices of the Royal Astronomical Society}
  \textbf{\bibinfo{volume}{384}}, \bibinfo{pages}{449–463}
  (\bibinfo{year}{2008}), ISSN \bibinfo{issn}{1365-2966},
  \urlprefix\url{http://dx.doi.org/10.1111/j.1365-2966.2007.12353.x}.

\bibitem[{\citenamefont{Feroz et~al.}(2009)\citenamefont{Feroz, Hobson, and
  Bridges}}]{MultiNest2}
\bibinfo{author}{\bibfnamefont{F.}~\bibnamefont{Feroz}},
  \bibinfo{author}{\bibfnamefont{M.~P.} \bibnamefont{Hobson}},
  \bibnamefont{and} \bibinfo{author}{\bibfnamefont{M.}~\bibnamefont{Bridges}},
  \bibinfo{journal}{Monthly Notices of the Royal Astronomical Society}
  \textbf{\bibinfo{volume}{398}}, \bibinfo{pages}{1601–1614}
  (\bibinfo{year}{2009}), ISSN \bibinfo{issn}{1365-2966},
  \urlprefix\url{http://dx.doi.org/10.1111/j.1365-2966.2009.14548.x}.

\bibitem[{\citenamefont{Feroz et~al.}(2019)\citenamefont{Feroz, Hobson,
  Cameron, and Pettitt}}]{MultiNest3}
\bibinfo{author}{\bibfnamefont{F.}~\bibnamefont{Feroz}},
  \bibinfo{author}{\bibfnamefont{M.~P.} \bibnamefont{Hobson}},
  \bibinfo{author}{\bibfnamefont{E.}~\bibnamefont{Cameron}}, \bibnamefont{and}
  \bibinfo{author}{\bibfnamefont{A.~N.} \bibnamefont{Pettitt}},
  \bibinfo{journal}{The Open Journal of Astrophysics}
  \textbf{\bibinfo{volume}{2}} (\bibinfo{year}{2019}), ISSN
  \bibinfo{issn}{2565-6120},
  \urlprefix\url{http://dx.doi.org/10.21105/astro.1306.2144}.

\bibitem[{\citenamefont{Buchner et~al.}(2014)\citenamefont{Buchner,
  Georgakakis, Nandra, Hsu, Rangel, Brightman, Merloni, Salvato, Donley, and
  Kocevski}}]{PyMultiNest}
\bibinfo{author}{\bibfnamefont{J.}~\bibnamefont{Buchner}},
  \bibinfo{author}{\bibfnamefont{A.}~\bibnamefont{Georgakakis}},
  \bibinfo{author}{\bibfnamefont{K.}~\bibnamefont{Nandra}},
  \bibinfo{author}{\bibfnamefont{L.}~\bibnamefont{Hsu}},
  \bibinfo{author}{\bibfnamefont{C.}~\bibnamefont{Rangel}},
  \bibinfo{author}{\bibfnamefont{M.}~\bibnamefont{Brightman}},
  \bibinfo{author}{\bibfnamefont{A.}~\bibnamefont{Merloni}},
  \bibinfo{author}{\bibfnamefont{M.}~\bibnamefont{Salvato}},
  \bibinfo{author}{\bibfnamefont{J.}~\bibnamefont{Donley}}, \bibnamefont{and}
  \bibinfo{author}{\bibfnamefont{D.}~\bibnamefont{Kocevski}},
  \bibinfo{journal}{Astronomy \& Astrophysics} \textbf{\bibinfo{volume}{564}},
  \bibinfo{pages}{A125} (\bibinfo{year}{2014}).

\bibitem[{\citenamefont{Lewis}(2019)}]{getdist}
\bibinfo{author}{\bibfnamefont{A.}~\bibnamefont{Lewis}} (\bibinfo{year}{2019}),
  \eprint{1910.13970}.

\bibitem[{\citenamefont{Hu and White}(1996)}]{Hu:1996}
\bibinfo{author}{\bibfnamefont{W.}~\bibnamefont{Hu}} \bibnamefont{and}
  \bibinfo{author}{\bibfnamefont{M.}~\bibnamefont{White}},
  \bibinfo{journal}{The Astrophysical Journal} \textbf{\bibinfo{volume}{471}},
  \bibinfo{pages}{30} (\bibinfo{year}{1996}).

\bibitem[{\citenamefont{{Cyr-Racine} et~al.}(2022)\citenamefont{{Cyr-Racine},
  {Ge}, and {Knox}}}]{cyr-racine2022}
\bibinfo{author}{\bibfnamefont{F.-Y.} \bibnamefont{{Cyr-Racine}}},
  \bibinfo{author}{\bibfnamefont{F.}~\bibnamefont{{Ge}}}, \bibnamefont{and}
  \bibinfo{author}{\bibfnamefont{L.}~\bibnamefont{{Knox}}},
  \bibinfo{journal}{\prl} \textbf{\bibinfo{volume}{128}}, \bibinfo{eid}{201301}
  (\bibinfo{year}{2022}), \eprint{2107.13000}.

\bibitem[{\citenamefont{Hou et~al.}(2013)\citenamefont{Hou, Keisler, Knox,
  Millea, and Reichardt}}]{Knox-neutrinos-damping-tail}
\bibinfo{author}{\bibfnamefont{Z.}~\bibnamefont{Hou}},
  \bibinfo{author}{\bibfnamefont{R.}~\bibnamefont{Keisler}},
  \bibinfo{author}{\bibfnamefont{L.}~\bibnamefont{Knox}},
  \bibinfo{author}{\bibfnamefont{M.}~\bibnamefont{Millea}}, \bibnamefont{and}
  \bibinfo{author}{\bibfnamefont{C.}~\bibnamefont{Reichardt}},
  \bibinfo{journal}{Physical Review D} \textbf{\bibinfo{volume}{87}}
  (\bibinfo{year}{2013}), ISSN \bibinfo{issn}{1550-2368},
  \urlprefix\url{http://dx.doi.org/10.1103/PhysRevD.87.083008}.

\bibitem[{\citenamefont{Bernal et~al.}(2016)\citenamefont{Bernal, Verde, and
  Riess}}]{trouble-h0}
\bibinfo{author}{\bibfnamefont{J.~L.} \bibnamefont{Bernal}},
  \bibinfo{author}{\bibfnamefont{L.}~\bibnamefont{Verde}}, \bibnamefont{and}
  \bibinfo{author}{\bibfnamefont{A.~G.} \bibnamefont{Riess}},
  \bibinfo{journal}{JCAP} \textbf{\bibinfo{volume}{10}}, \bibinfo{pages}{019}
  (\bibinfo{year}{2016}), \eprint{1607.05617}.

\bibitem[{PLA()}]{PLA}
\emph{\bibinfo{title}{Planck legacy archive}},
  \bibinfo{howpublished}{\url{https://pla.esac.esa.int/\#cosmology}}.

\bibitem[{\citenamefont{Riess et~al.}(2018)}]{Riess18}
\bibinfo{author}{\bibfnamefont{A.~G.} \bibnamefont{Riess}}
  \bibnamefont{et~al.}, \bibinfo{journal}{Astrophys. J.}
  \textbf{\bibinfo{volume}{861}}, \bibinfo{pages}{126} (\bibinfo{year}{2018}),
  \eprint{1804.10655}.

\bibitem[{\citenamefont{Ade et~al.}(2016)}]{PCP15}
\bibinfo{author}{\bibfnamefont{P.~A.~R.} \bibnamefont{Ade}}
  \bibnamefont{et~al.} (\bibinfo{collaboration}{Planck}),
  \bibinfo{journal}{Astron. Astrophys.} \textbf{\bibinfo{volume}{594}},
  \bibinfo{pages}{A13} (\bibinfo{year}{2016}), \eprint{1502.01589}.

\bibitem[{\citenamefont{Clark et~al.}(2021)\citenamefont{Clark, Vattis, Fan,
  and Koushiappas}}]{Clark:2021hlo}
\bibinfo{author}{\bibfnamefont{S.~J.} \bibnamefont{Clark}},
  \bibinfo{author}{\bibfnamefont{K.}~\bibnamefont{Vattis}},
  \bibinfo{author}{\bibfnamefont{J.}~\bibnamefont{Fan}}, \bibnamefont{and}
  \bibinfo{author}{\bibfnamefont{S.~M.} \bibnamefont{Koushiappas}}
  (\bibinfo{year}{2021}), \eprint{2110.09562}.

\bibitem[{\citenamefont{Smith et~al.}(2021)\citenamefont{Smith, Poulin, Bernal,
  Boddy, Kamionkowski, and Murgia}}]{EDE-prior-volume}
\bibinfo{author}{\bibfnamefont{T.~L.} \bibnamefont{Smith}},
  \bibinfo{author}{\bibfnamefont{V.}~\bibnamefont{Poulin}},
  \bibinfo{author}{\bibfnamefont{J.~L.} \bibnamefont{Bernal}},
  \bibinfo{author}{\bibfnamefont{K.~K.} \bibnamefont{Boddy}},
  \bibinfo{author}{\bibfnamefont{M.}~\bibnamefont{Kamionkowski}},
  \bibnamefont{and} \bibinfo{author}{\bibfnamefont{R.}~\bibnamefont{Murgia}},
  \bibinfo{journal}{Physical Review D} \textbf{\bibinfo{volume}{103}}
  (\bibinfo{year}{2021}), ISSN \bibinfo{issn}{2470-0029},
  \urlprefix\url{http://dx.doi.org/10.1103/PhysRevD.103.123542}.

\bibitem[{\citenamefont{Herold et~al.}(2021)\citenamefont{Herold, Ferreira, and
  Komatsu}}]{EDE-laura-likelihood}
\bibinfo{author}{\bibfnamefont{L.}~\bibnamefont{Herold}},
  \bibinfo{author}{\bibfnamefont{E.~G.~M.} \bibnamefont{Ferreira}},
  \bibnamefont{and} \bibinfo{author}{\bibfnamefont{E.}~\bibnamefont{Komatsu}}
  (\bibinfo{year}{2021}), \eprint{2112.12140}.

\bibitem[{\citenamefont{Efstathiou}(2021)}]{ToH0OrNotToH0}
\bibinfo{author}{\bibfnamefont{G.}~\bibnamefont{Efstathiou}},
  \bibinfo{journal}{Monthly Notices of the Royal Astronomical Society}
  \textbf{\bibinfo{volume}{505}}, \bibinfo{pages}{3866} (\bibinfo{year}{2021}).

\bibitem[{\citenamefont{Camarena and Marra}(2021)}]{Camarena-21}
\bibinfo{author}{\bibfnamefont{D.}~\bibnamefont{Camarena}} \bibnamefont{and}
  \bibinfo{author}{\bibfnamefont{V.}~\bibnamefont{Marra}},
  \bibinfo{journal}{Mon. Not. Roy. Astron. Soc.}
  \textbf{\bibinfo{volume}{504}}, \bibinfo{pages}{5164} (\bibinfo{year}{2021}),
  \eprint{2101.08641}.

\bibitem[{\citenamefont{Benevento et~al.}(2020)\citenamefont{Benevento, Hu, and
  Raveri}}]{HuBenevento:2020}
\bibinfo{author}{\bibfnamefont{G.}~\bibnamefont{Benevento}},
  \bibinfo{author}{\bibfnamefont{W.}~\bibnamefont{Hu}}, \bibnamefont{and}
  \bibinfo{author}{\bibfnamefont{M.}~\bibnamefont{Raveri}},
  \bibinfo{journal}{Phys. Rev. D} \textbf{\bibinfo{volume}{101}},
  \bibinfo{pages}{103517} (\bibinfo{year}{2020}),
  \urlprefix\url{https://link.aps.org/doi/10.1103/PhysRevD.101.103517}.

\bibitem[{\citenamefont{Alestas and Perivolaropoulos}(2021)}]{Alestas:2021}
\bibinfo{author}{\bibfnamefont{G.}~\bibnamefont{Alestas}} \bibnamefont{and}
  \bibinfo{author}{\bibfnamefont{L.}~\bibnamefont{Perivolaropoulos}},
  \bibinfo{journal}{Monthly Notices of the Royal Astronomical Society}
  \textbf{\bibinfo{volume}{504}}, \bibinfo{pages}{3956} (\bibinfo{year}{2021}),
  ISSN \bibinfo{issn}{0035-8711},
  \eprint{https://academic.oup.com/mnras/article-pdf/504/3/3956/37906343/stab1070.pdf},
  \urlprefix\url{https://doi.org/10.1093/mnras/stab1070}.

\bibitem[{\citenamefont{Yang et~al.}(2020)\citenamefont{Yang, Di Valentino,
  Pan, Wu, and Lu}}]{Weiqiang-Valentino:2020}
\bibinfo{author}{\bibfnamefont{W.}~\bibnamefont{Yang}},
  \bibinfo{author}{\bibfnamefont{E.}~\bibnamefont{Di Valentino}},
  \bibinfo{author}{\bibfnamefont{S.}~\bibnamefont{Pan}},
  \bibinfo{author}{\bibfnamefont{Y.}~\bibnamefont{Wu}}, \bibnamefont{and}
  \bibinfo{author}{\bibfnamefont{J.}~\bibnamefont{Lu}},
  \bibinfo{journal}{Monthly Notices of the Royal Astronomical Society}
  \textbf{\bibinfo{volume}{501}}, \bibinfo{pages}{5845} (\bibinfo{year}{2020}),
  ISSN \bibinfo{issn}{0035-8711},
  \eprint{https://academic.oup.com/mnras/article-pdf/501/4/5845/36053888/staa3914.pdf},
  \urlprefix\url{https://doi.org/10.1093/mnras/staa3914}.

\bibitem[{\citenamefont{Alestas et~al.}(2022)\citenamefont{Alestas, Camarena,
  Di~Valentino, Kazantzidis, Marra, Nesseris, and
  Perivolaropoulos}}]{Alestas:2021luu}
\bibinfo{author}{\bibfnamefont{G.}~\bibnamefont{Alestas}},
  \bibinfo{author}{\bibfnamefont{D.}~\bibnamefont{Camarena}},
  \bibinfo{author}{\bibfnamefont{E.}~\bibnamefont{Di~Valentino}},
  \bibinfo{author}{\bibfnamefont{L.}~\bibnamefont{Kazantzidis}},
  \bibinfo{author}{\bibfnamefont{V.}~\bibnamefont{Marra}},
  \bibinfo{author}{\bibfnamefont{S.}~\bibnamefont{Nesseris}}, \bibnamefont{and}
  \bibinfo{author}{\bibfnamefont{L.}~\bibnamefont{Perivolaropoulos}},
  \bibinfo{journal}{Phys. Rev. D} \textbf{\bibinfo{volume}{105}},
  \bibinfo{pages}{063538} (\bibinfo{year}{2022}), \eprint{2110.04336}.

\bibitem[{\citenamefont{Nunes and Di~Valentino}(2021)}]{Nunes:2021zzi}
\bibinfo{author}{\bibfnamefont{R.~C.} \bibnamefont{Nunes}} \bibnamefont{and}
  \bibinfo{author}{\bibfnamefont{E.}~\bibnamefont{Di~Valentino}},
  \bibinfo{journal}{Phys. Rev. D} \textbf{\bibinfo{volume}{104}},
  \bibinfo{pages}{063529} (\bibinfo{year}{2021}), \eprint{2107.09151}.

\bibitem[{\citenamefont{Sekiguchi and Takahashi}(2021)}]{Sekiguchi:2020teg}
\bibinfo{author}{\bibfnamefont{T.}~\bibnamefont{Sekiguchi}} \bibnamefont{and}
  \bibinfo{author}{\bibfnamefont{T.}~\bibnamefont{Takahashi}},
  \bibinfo{journal}{Phys. Rev. D} \textbf{\bibinfo{volume}{103}},
  \bibinfo{pages}{083507} (\bibinfo{year}{2021}), \eprint{2007.03381}.

\bibitem[{\citenamefont{Vagnozzi et~al.}(2021)\citenamefont{Vagnozzi, Pacucci,
  and Loeb}}]{Vagnozzi:2021tjv}
\bibinfo{author}{\bibfnamefont{S.}~\bibnamefont{Vagnozzi}},
  \bibinfo{author}{\bibfnamefont{F.}~\bibnamefont{Pacucci}}, \bibnamefont{and}
  \bibinfo{author}{\bibfnamefont{A.}~\bibnamefont{Loeb}}
  (\bibinfo{year}{2021}), \eprint{2105.10421}.

\bibitem[{\citenamefont{Kumar and Nunes}(2016)}]{Kumar:2016zpg}
\bibinfo{author}{\bibfnamefont{S.}~\bibnamefont{Kumar}} \bibnamefont{and}
  \bibinfo{author}{\bibfnamefont{R.~C.} \bibnamefont{Nunes}},
  \bibinfo{journal}{Phys. Rev. D} \textbf{\bibinfo{volume}{94}},
  \bibinfo{pages}{123511} (\bibinfo{year}{2016}), \eprint{1608.02454}.

\bibitem[{\citenamefont{Lucca}(2021)}]{Lucca:2021eqy}
\bibinfo{author}{\bibfnamefont{M.}~\bibnamefont{Lucca}},
  \bibinfo{journal}{Phys. Rev. D} \textbf{\bibinfo{volume}{104}},
  \bibinfo{pages}{083510} (\bibinfo{year}{2021}), \eprint{2106.15196}.

\bibitem[{\citenamefont{G\'omez-Valent
  et~al.}(2020)\citenamefont{G\'omez-Valent, Pettorino, and
  Amendola}}]{Gomez:Coupled-DE-H0}
\bibinfo{author}{\bibfnamefont{A.}~\bibnamefont{G\'omez-Valent}},
  \bibinfo{author}{\bibfnamefont{V.}~\bibnamefont{Pettorino}},
  \bibnamefont{and} \bibinfo{author}{\bibfnamefont{L.}~\bibnamefont{Amendola}},
  \bibinfo{journal}{Phys. Rev. D} \textbf{\bibinfo{volume}{101}},
  \bibinfo{pages}{123513} (\bibinfo{year}{2020}), \eprint{2004.00610}.

\bibitem[{\citenamefont{Di~Valentino}(2021)}]{Valentino2021combined}
\bibinfo{author}{\bibfnamefont{E.}~\bibnamefont{Di~Valentino}},
  \bibinfo{journal}{Mon. Not. Roy. Astron. Soc.}
  \textbf{\bibinfo{volume}{502}}, \bibinfo{pages}{2065} (\bibinfo{year}{2021}),
  \eprint{2011.00246}.

\bibitem[{\citenamefont{Armendariz-Picon
  et~al.}(2001)\citenamefont{Armendariz-Picon, Mukhanov, and
  Steinhardt}}]{essentials-of-k-essence-picon-mukhanov-steinhardt}
\bibinfo{author}{\bibfnamefont{C.}~\bibnamefont{Armendariz-Picon}},
  \bibinfo{author}{\bibfnamefont{V.}~\bibnamefont{Mukhanov}}, \bibnamefont{and}
  \bibinfo{author}{\bibfnamefont{P.~J.} \bibnamefont{Steinhardt}},
  \bibinfo{journal}{Phys. Rev. D} \textbf{\bibinfo{volume}{63}},
  \bibinfo{pages}{103510} (\bibinfo{year}{2001}),
  \urlprefix\url{https://link.aps.org/doi/10.1103/PhysRevD.63.103510}.

\bibitem[{\citenamefont{DeDeo et~al.}(2003)\citenamefont{DeDeo, Caldwell, and
  Steinhardt}}]{DeDeo-Caldwell-Steinhardt-k-essence}
\bibinfo{author}{\bibfnamefont{S.}~\bibnamefont{DeDeo}},
  \bibinfo{author}{\bibfnamefont{R.~R.} \bibnamefont{Caldwell}},
  \bibnamefont{and} \bibinfo{author}{\bibfnamefont{P.~J.}
  \bibnamefont{Steinhardt}}, \bibinfo{journal}{Phys. Rev. D}
  \textbf{\bibinfo{volume}{67}}, \bibinfo{pages}{103509}
  (\bibinfo{year}{2003}), \bibinfo{note}{[Erratum: Phys.Rev.D 69, 129902
  (2004)]}, \eprint{astro-ph/0301284}.

\bibitem[{\citenamefont{Miyatake et~al.}(2021)}]{HSC_2021}
\bibinfo{author}{\bibfnamefont{H.}~\bibnamefont{Miyatake}} \bibnamefont{et~al.}
  (\bibinfo{year}{2021}), \eprint{2111.02419}.

\end{thebibliography}

\newpage

\appendix

\section{Brane Model}
\label{sec:A}
Here we study a related quintessence potential, dubbed the ``Brane'' model \cite{Skordis_2002}, given by,
\begin{equation}
    V=\left[\frac{C}{(\phi-B)^{2}+A}+D\right] e^{-\lambda \phi}    
    \label{eq:Brane}
\end{equation}
Our purpose is not to extensively study the details of the parameter space of this model, though they are quite similar to the AS case, but to demonstrate the point that the oft-studied pure ``tracking'' quintessence potentials are {\it a-priori} unfit for addressing the $H_0$ tension. To this end, we set $C=D=0.01$, and $A=0.0025$. In order get the best-fit parameters, we set $\lambda=10.5$ and find the corresponding value of $B, \phi_{\rm ini}$, and $\phi'_{\rm ini}$ using the same procedure as in Section 4.2. 

As with the AS potential, for suitable choices of the free parameters, the Brane model is dominated by the pure exponential factor far from the minimum. However, unlike the AS case, the Brane field tracks the dominant background component during {\it both} the radiation and the matter-dominated regime, as shown in Fig.~\ref{fig:brane-as-bf} (left-panel). This is precisely why it fails to increase $H_0$ above the $\Lambda$CDM value: $\omega_{cdm}$ must increase to compensate for the change in the radiation driving envelope induced by the quintessence component during radiation domination (see Section 5), as in the AS or $N_{\rm eff}$ case, which inevitably increases the expansion rate post-recombination. However, unlike the AS and $N_{\rm eff}$ cases, there is now an additional significant contribution to the expansion rate of $\mathcal{O}(3/\lambda^2)$ from the matter-era tracking of the Brane potential. Thus, even though the pre-recombination contribution from the Brane field is able to decrease the sound horizon scale by approximately the same amount as the AS case, the increased expansion rate in the 2 decades of scale factor post-recombination overshoots the requisite change in $D_M^*$ such that $H(z\rightarrow 0)$ must in fact {\it decrease} (relative to $\Lambda$CDM) in order to compensate for this overshooting, as shown in Fig.~\ref{fig:brane-as-bf} (right panel). Moreover, the contribution of the Brane field during matter-domination causes an even further departure from the $H(z)\rightarrow \alpha H(z)$ scaling solution (discussed in Section 5) so that the fit to the Planck and BAO data is abysmal at $\Delta \chi^2 \approx 100$. Finally, notice that there is even more power suppression in the Brane case since modes are now suppressed throughout the matter dominated regime, as opposed to the scale-dependent suppression in the ASCDM model (c.f. Fig~\ref{fig:mpk}). We summarize the various parameters of interest for the best-fit case with {\bf fixed $\lambda=10.5$} in Table~\ref{tab:brane-param-table}. 

Worth noting is that the Brane field approached $w_\phi \approx -1$ today independently of the choice of $\lambda$ (as long as the potential admits a false vacuum; see point 1 in Section 4.2). This is in contrast to the oft-used Ratra-Peebles type potentials  \cite{RP1988, Ooba-Ratra-Sugiyama} where the field is ``frozen'' by Hubble friction and $w_\phi$ depends on the dynamical parameter so that the physical interpretation of parameter inference results that disfavor such models may be ambiguous. The Brane model can resolve this ambiguity: the fact that \texttt{Planck+BAO} disfavor the Brane potential (relative to $\Lambda$CDM) shows an aversion strictly to tracking fields that contribute significantly during matter domination.  

\begin{figure*}
    \centering
    \subfigure{\includegraphics[clip,width=\columnwidth]{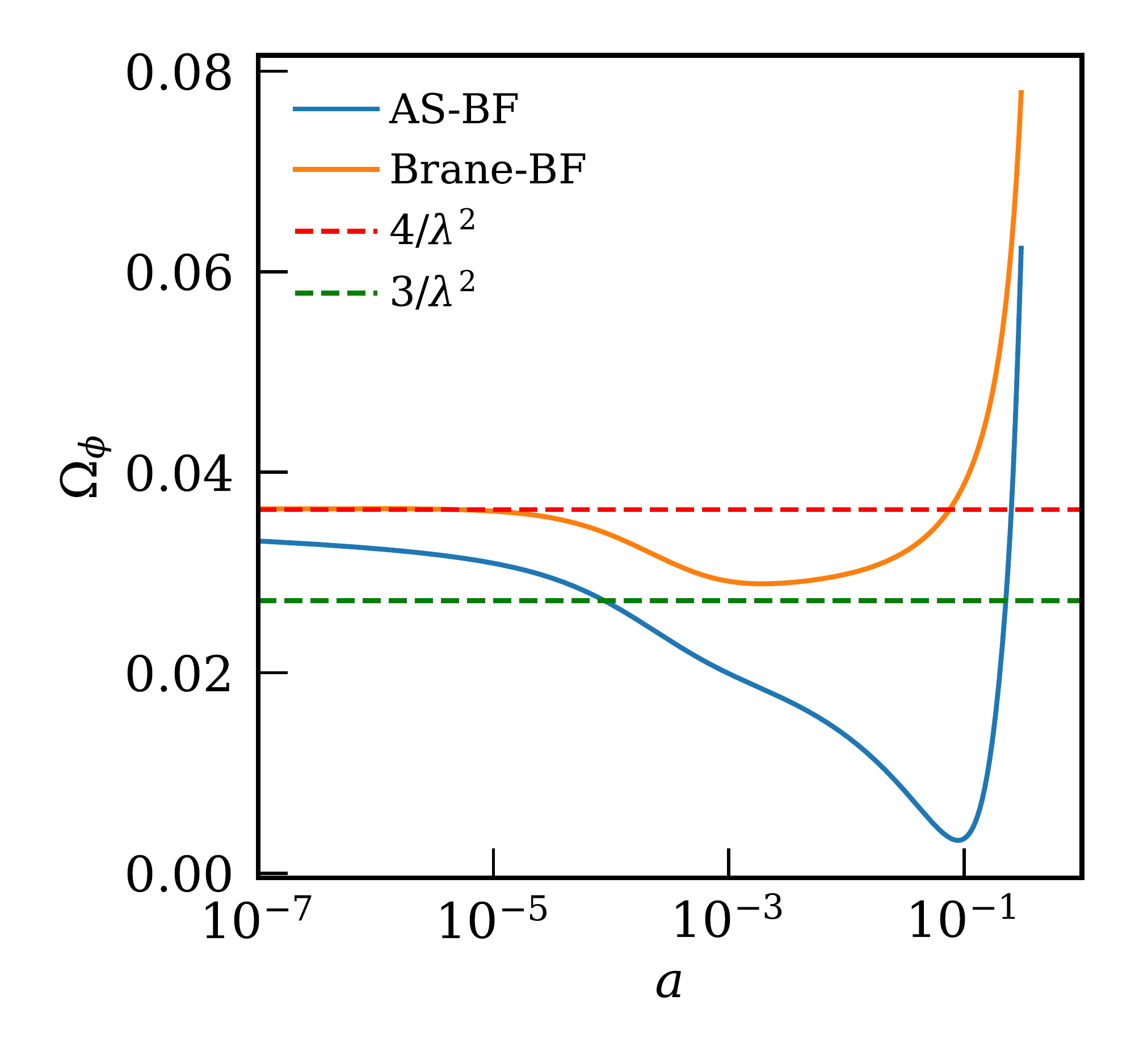}}\quad
    \subfigure{\includegraphics[clip,width=\columnwidth]{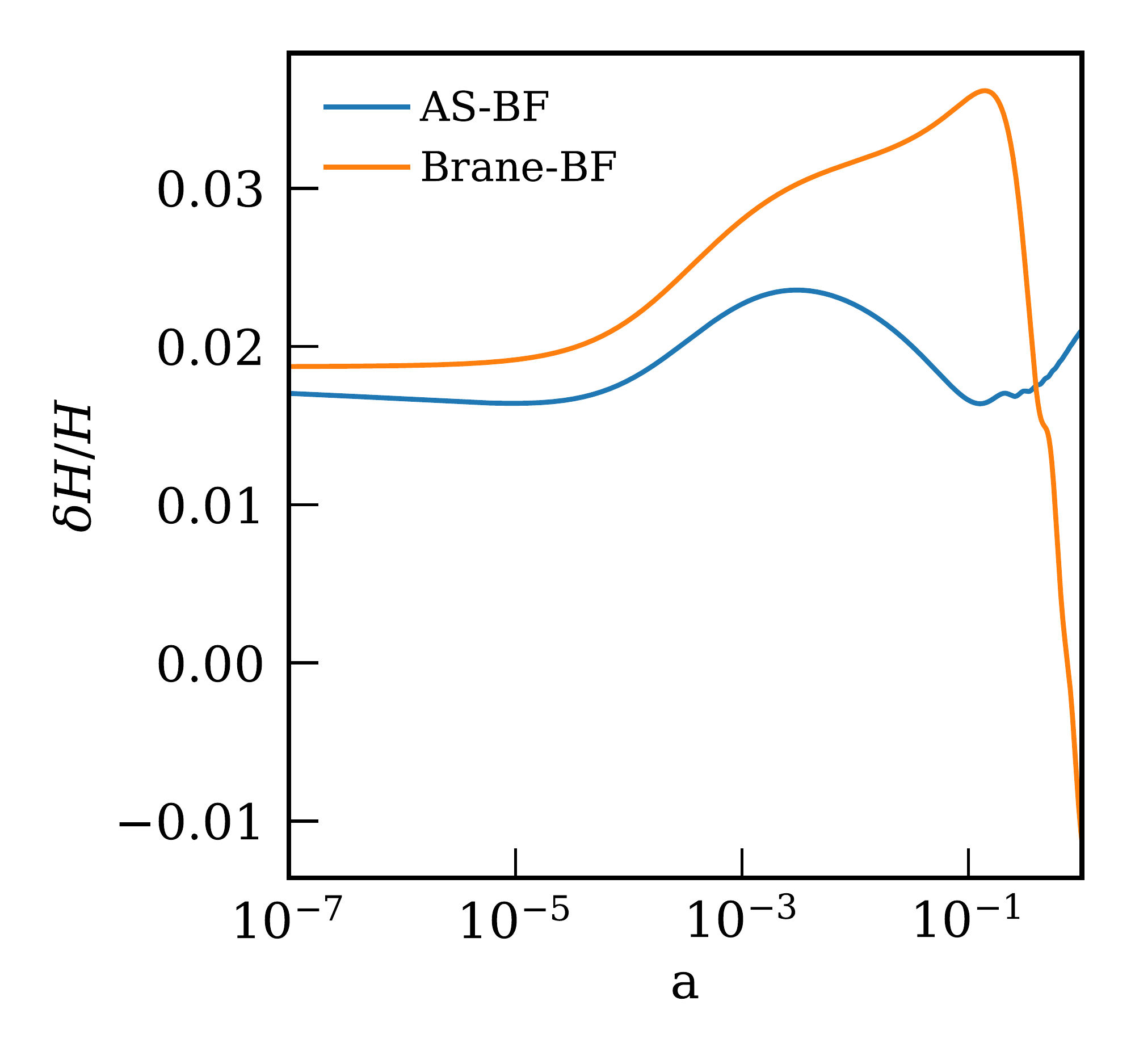}}

    \caption{The figure depicts the behavior of the Brane quintessence field, along with the AS potential for comparison, for the best-fit parameters with fixed $\lambda=10.5$ (c.f. Fig~\ref{fig:As-Neff-scaling}). The left panel shows the contribution to the total energy density from the quintessence fields. Notice that the Brane field approaches the scaling attractor during matter domination while the AS field breaks away from it. The right panel shows the fractional change in the expansion rate with respect to the best-fit $\Lambda$CDM model; the Brane field is unable to increase $H_0$ even for a sizeable pre-recombination contribution.}
    \label{fig:brane-as-bf}
\end{figure*}

\begin{table}[htb]
\begin{ruledtabular}
\begin{tabular}{p{2cm}|c |c |c}
&Brane & AS & \lcdm \\
\hline
$\lambda$ & 10.5 & 10.5 & \\
$\omega_{\rm cdm}$ & 0.124 & 0.124 & 0.120 \\
$H_0$[km ${\rm s}^{-1} {\rm Mpc}^{-1}$] &66.81 & 68.99&67.57\\
$r_s^*$[Mpc] &141.3 &141.9 &144.9\\
$\sigma_8$ & 0.709&0.782 &0.805\\
$S_8$ & 0.744 &0.794 &0.822\\
\end{tabular}
\end{ruledtabular}
\label{tab:brane-param-table}
\caption{Best-fit parameters for $\lambda=10.5$ fixed for the AS and Brane cases using \texttt{Planck} temperature and polarization, and BOSS BAO (DR12) data. Clearly, though $\omega_{\rm cdm}$ and $r_s^*$ change by the same amount in both the quintessence models, the matter dominated tracking solution in the Brane case leads to a decreased inference of both $H_0$ and $S_8$ compared to the AS case.}

\end{table}

\end{document}